\documentclass[floatfix,aps,showpacs,twocolumn,amsmath,amssymb,superscriptaddress]{revtex4}  %
\usepackage{amsmath}
\usepackage{graphicx}
\usepackage{dcolumn}
\usepackage{amsmath}
\usepackage{subfigure,hyperref}
\usepackage[normalem,normalbf]{ulem}
\usepackage{enumerate}
\usepackage{tikz}
\usepackage{algpseudocode}
\usepackage{algorithm}
\usepackage{listings}
\usepackage{color}
\usepackage{mleftright}
\usepackage{natbib}
\usepackage{accents}
\usepackage{stackengine}
\usepackage{placeins}
\usepackage[english]{babel}

\usepackage{scalerel,stackengine}
\stackMath
\newcommand\reallywidehat[1]{%
\savestack{\tmpbox}{\stretchto{%
  \scaleto{%
    \scalerel*[\widthof{\ensuremath{#1}}]{\kern-.6pt\bigwedge\kern-.6pt}%
    {\rule[-\textheight/2]{1ex}{\textheight}}%WIDTH-LIMITED BIG WEDGE
  }{\textheight}% 
}{0.5ex}}%
\stackon[1pt]{#1}{\tmpbox}%
}

\DeclareMathOperator*{\argmin}{arg\,min}

\newcommand{\subsubsubsection}[1]{\paragraph{#1}\mbox{}\\}
\definecolor{dkgreen}{rgb}{0,0.6,0}
\definecolor{gray}{rgb}{0.5,0.5,0.5}
\definecolor{mauve}{rgb}{0.58,0,0.82}
\newcommand{\bc}{\begin{center}}
\newcommand{\ec}{\end{center}}
\newcommand{\be}{\begin{equation}}
\newcommand{\ee}{\end{equation}}
\newcommand{\bea}{\begin{eqnarray}}
\newcommand{\eea}{\end{eqnarray}}
\newcommand{\beq}{\begin{eqnarray*}}
\newcommand{\eeq}{\end{eqnarray*}}
\newcommand{\bv}{\left( \begin{array}{c} }
\newcommand{\ev}{\end{array} \right) }

\newcommand{\mean}{\mathrm{mean}}
\newcommand{\std}{\mathrm{std}}

\newcommand{\ve}[1]{\boldsymbol{#1}}

\usetikzlibrary{arrows,chains,matrix,positioning,scopes}

\makeatletter
\tikzset{join/.code=\tikzset{after node path={\ifx\tikzchainprevious\pgfutil@empty\else(\tikzchainprevious) edge[every join]#1(\tikzchaincurrent)\fi}}}
\makeatother
\tikzset{>=stealth',every on chain/.append style={join},every join/.style={->}}
\tikzstyle{labeled}=[execute at begin node=$\scriptstyle,execute at end node=$]

\usepackage{enumerate}

\begin{document}

\title{Systematic Asset Allocation using Flexible Views for South African Markets}

\author{Ann Sebastian}
\affiliation{Department of Statistical Science, University of Cape Town, Rondebosch, South Africa}
\affiliation{Index Investments, STANLIB Asset Management (Pty) Ltd, Johannesburg, South Africa}
\email{ann.sebastian@stanlib.com}
\author{Tim Gebbie}
\email{tim.gebbie@uct.ac.za}
\affiliation{Department of Statistical Science, University of Cape Town, Rondebosch, South Africa}
%\author{??}
%
\begin{abstract}
We implement a systematic asset allocation model using the Historical Simulation with Flexible Probabilities (HS-FP) framework developed by Meucci \cite{A2010,A2012,A2013}. The HS-FP framework is a flexible non-parametric estimation approach that considers future asset class behavior to be conditional on time and market environments, and derives a forward looking distribution that is consistent with this view while remaining close as possible to the prior distribution. The framework derives the forward looking distribution by applying unequal time and state conditioned probabilities to historical observations of asset class returns. This is achieved using relative entropy to find estimates with the least distortion to the prior distribution. Here, we use the HS-FP framework on South African financial market data for asset allocation purposes; by estimating expected returns, correlations and volatilities that are better represented through the measured market cycle. We demonstrated a range of state variables that can be useful towards understanding market environments. Concretely, we compare the out-of-sample performance for a specific configuration of the HS-FP model relative to classic Mean Variance Optimization(MVO) and Equally Weighted (EW) benchmark models. The framework displays low probability of backtest overfitting and the out-of-sample net returns and Sharpe ratio point estimates of the HS-FP model outperforms the benchmark models. However, the results are inconsistent when training windows are varied, the Sharpe ratio is seen to be inflated, and the method does not demonstrate statistically significant out-performance on a gross and net basis
\end{abstract}

\maketitle
%\tableofcontents
\section{Introduction}

Asset allocation is the process of deciding how much to allocate to asset classes by gathering and processing market related data in order to deliver on client objectives \cite{B2000}. Asset allocation is considered the most important determinant of portfolio outcomes \cite{A1986} not only because asset level portfolio constraints dominate portfolio construction but also because it is generally the most difficult decision for investors to make. For investors to decide how much to allocate to various asset classes they need to gather views on each of the asset classes. Asset class views are notoriously correlated, dynamic, inter-related and noisy. A systematic approach to asset allocation brings discipline by implementing a rules based approach of including relevant information about the market in the return prospect of asset classes \cite{B1983}. The key inputs required in any quantitative portfolio construction process grounded in modern portfolio theory are the underlying market risk drivers, return views, and views on the risk and dependence structure of the asset class \cite{C2000}. 

However, these inputs are unknown and need to be estimated by market participants. Market participants can estimate these asset characteristics from historical data or simulated data. Here we follow a data-informed approach and use historical data. Some of the benefits of using historical data include that it is simple to implement and represents the actual asset class history so assumptions on asset class distribution, dependence structure and other features need not be explicitly made. But, naively using measured historical data of financial markets presupposes that past experience will somehow match future experience. This is unlikely.

The reality is that we only have a single measured realization of reality but do not explicitly know the probability of the sequence of history we have measured, we neither have a concrete measure of how quickly information dissipates, how agents and market participants adapt to the data they measure, nor how quickly our models can be expected to break-down. This opens all modeling using measured financial market history to potentially wild forms of generalisation errors.

Although there has been an abundance in academic literature supporting asset class return predictability, as summarised by Rapach and Zhou in \cite{B2013}, and there is general consensus, that in-sample asset class returns do display some sort of predictability \cite{A2000}; there is far less consensus regarding out-of-sample asset class return predictability \cite{A2008}. 
  
For these reasons a systematic asset allocation model is required that is adaptive to changing market conditions and can incorporate historical data with investor views in a mathematically robust manner. Investors typically trust recent history more than those from far in the past. With this in mind we considered the ``Historical Simulation with Flexible Probabilities'' (HS-FP) framework developed by Meucci \cite{A2010,A2012,A2013}. To effectively adapt to changing market conditions, the framework needs to increase the breadth of data while reducing its dependency on data depth. 

The key features of the HS-FP framework are that it is, first, a non-parametric approach, and second, it is able to incorporate historical data with investor views on state variables. This is achieved by generating time and state conditioned probabilities. The type of investor view we make use of relates to the qualitative question investors often ask: 
{\it When, in recent history, were state variables most similar to today’s levels?}
We aim to estimate empirical distributions which reflect the current market regime. This predictive requirement can be implemented using the most recent mixed distribution estimate to form subsequent period return forecast. The paper will also consider using state variables for asset class return predictability. 

When asset class returns are observed through a market cycle it can be seen that the returns vary through time and market regimes \cite{A2007}. Different asset classes perform differently in different market conditions and no single asset class dominates in all market conditions. Additionally, state variables exist that are useful in distinguishing market conditions and therefore forecasting asset class performance. State variables are observable indicators that measure the state of the market. There is a variety of state variable indicators ranging from macro economic, financial, risk and other categories. Flavin and Wickens \cite{A2003} provide a detailed overview of key previous literature on using macro economic and financial variables to forecast asset class returns. Here we restrict ourselves to a reduced subset of explicable state variables - this need not be the case. 

The key questions clients often ask about any investment or trading idea is: {\it Can the idea deliver in a real way?} We could naively argue, {\it Yes}, based on a single out-of-sample cumulative return time series from a single historical simulation compared to judiciously chosen benchmarks (see our Figure \ref{fig:fig9}), however, this would be misleading. A client would usually follow the value-add question by the risk-qualifier question: {\it Can this protect me against bad events?} That is almost impossible to effectively answer. 

The general client approach to investments is to often ignore strategy details until there is some sort of problem, and to then expect a concrete explanation or an easily understandable response. Using explicable state variables with a concrete economic back-story has a client facing advantage even if one is not yet using all the cross-sectional information currently available. Technology that is sold by arguing that it can offer better investment or trading decisions is often regarded and evaluated differently from the same technology being sold to the client on the basis that it can reduce costs, increase efficiencies and reduce implementation risk. Having a story still matters. There is still an ease of client buy-in when faced with complex technologies that cannot provide a clear reason for specific decisions, particularly when they are of a data-informed and statistical nature. In asset management we are now clearly moving from the post-quant world into the world of unapologetically data-informed decision making. It will be increasingly difficult to move back into the world where it was relatively straight-forward to provide any explanation or story for all the processes we now deploy in asset management and trading. Reflection on systematic processes that are quantitative but still grounded in well understood explicable state variables can be a valuable bridge when facing clients even when the results may be statistically questionable.

For these reasons we think that the HS-FP approach is both interesting and useful as it can act as a conceptual bridge as we leap from the quant world (which is on the boundary of plausible econometric explanations for investment decision making) to that of pure data-informed decision making where statistical features are used on large scale to define state variables, but where there may be little hope for clear explanations and we are left to only rely on our estimation of generalization errors.

In the current work, we return to the classical problem of historical simulation and try to qualify the presentation of single out-of-sample cumulative return time series, as in our Figure \ref{fig:fig9}, to claim that the approach can add value in the recent past even after transaction costs in the excess of 50bps. But, here the key qualifier is that we cannot provide statistical certainty because we do not (and cannot feasibly) have long enough time-series to provide reasonable statistical significance or consistency under simulation, even when we have a low probability of back test over-fitting and a plausible back-story. 

We only have a historical simulation grounded in explicable state variables that shows that the strategy performed well (out-of-sample) during previous times in the past, when bad things happened ({\it e.g.} during the Global Financial Crises) and that the strategy had historical advantages over other investment choices over the recent past (say, the last 5 years). 

Concretely, the typical historical approach to asset class return predictability is to use all historical data equally to forecast future asset class returns. Here, the key step is to deviate from this, and apply unequal time and state variable conditioned probabilities to historical returns. 

How can this be tested through simulation in a way that can allow us to trust the approach? To draw conclusion about the efficacy of this approach to asset allocation we explicitly compare the HS-FP approach to the classic MVO approach \cite{A1952} and a naive EW approach observing the out-of-sample returns and other descriptive statistics. We then consider the statistical significance of these results and assess the potential of backtest overfitting. 

The objective of this work is to:
\begin{enumerate}
\item Demonstrate the use of the HS-FP model on South African financial market data for systematic asset allocation purposes;
\item Display a range of explicable state variables that can be used to understand market conditions;
\item Evaluate whether this time and state conditioned risk and return of asset classes provides any significant improvement on portfolio performance when comparing the out of sample performance of the HS-FP model to benchmark models.
\end{enumerate}
Although the key contribution of this paper is to consider the potential effectiveness of a simple HS-FP implementation in the context of South African markets, the work can be of interest to a more general audience as it links the HS-FP framework to an explicit benchmark based test, and a quantitative reflection on the perils of backtest overfitting as we step into the post-quant world of data-informed asset management aimed at making better decision and not merely reducing costs. 

The paper is organized as follows. In Section \ref{sec:hs-fp} we discuss the HS-FP model setup. In Section \ref{sec:bench} we cover the benchmark models used, in Section \ref{sec:implement} we outline the implementation, in Section \ref{sec:analysis} we analyze the out-of-sample results, and in Section \ref{sec:stats} we discuss whether these results can be considered sufficiently robust for our implementation to be practically useful for asset management use-cases. Finally, in section \ref{sec:conclude} we have the summary and conclusions.

In brief, we are able to show that the out-of-sample net returns and risk adjusted returns of HS-FP model outperforms the benchmark models. However, these results are inconsistent when training windows are varied and are not statistically significant. The point estimate Sharpe Ratio is inflated and the track record length fails to be above the minimum track record length required for statistical significance. The resulting view, although we are able to show a low probability that the backtested HS-FP model is over-fitting the data, the overall result indicates that the proposed approach cannot be used in isolation from other informed approaches to asset allocation. Although we think this is an important step towards algorithmic asset management, the process in itself has to be viewed as insufficient without informed qualitative over-sight, or process integration with alternative information sources to increase data width to address window dependencies and a general lack of depth in the measured data. 

\section{Historical Simulation with Flexible Probabilities Model Setup} \label{sec:hs-fp}

This section explains the HS-FP model setup. We detail the HS-FP framework developed by Meucci \cite{A2010,A2012,A2013} and specifically how this framework can be used in the estimation step of the investment process. In the following subsections we outline the involved techniques and how these techniques assist in forming the HS-FP framework. We start by showing how to use flexible probabilities to condition a single state variable in sub-section \ref{sec:fp-single} and then show how to use flexible probabilities to condition multiple state variables in sub-section \ref{sec:fp-mult}. 

\subsection{Flexible Probabilities} \label{sec:fp-single}

The HS-FP model commences at a generic time t, with a set of $\bar{n}$ joint market invariants $\boldsymbol{\varepsilon}_{t} \equiv (\varepsilon_{1,t},\hdots,\varepsilon_{\Bar{n},t})$ that are approximately independently and identically distributed (i.i.d.) across time, and considers a historical time series of the $\bar{n}$ market invariants $\{\boldsymbol{\epsilon}_{t} \equiv (\epsilon_{1,t},\hdots,\epsilon_{\Bar{n},t})\}^{\Bar{t}}_{t=1}$ where t=1 is the first and t=$\bar{t}$ is the most recent historical observation. If the number of historical observations in the time series of market invariants is large then using the law of large numbers the distribution of the future market invariant ${\varepsilon}_{\bar{t}+1}$ can be estimated non-parametrically from its historical time series $\{\boldsymbol{\epsilon}_{t} \equiv (\epsilon_{1,t},\hdots,\epsilon_{\Bar{n},t})\}^{\Bar{t}}_{t=1}$ \cite{Vap1971}.

The historical non-parametric approach to estimating future invariant distribution at current time $\bar{t}$, uses historical observations of the market invariant as forward looking scenarios with probabilities $\{p_t\}_{t=1}^{\bar{t}}$assigned to each of these historical scenarios such that the future distribution of market invariant is estimated as:

\begin{equation} \label{eq:1}
    \ve{\varepsilon}_{\bar{t}+1} \equiv 
    	\begin{pmatrix}
    		\varepsilon_{1,\bar{t}+1} \\
    		\vdots \\
    		\varepsilon_{\bar{n},\bar{t}+1}
    	\end{pmatrix}
   \sim
   \left\{ { \ve{\epsilon}_t \equiv 
   \begin{pmatrix}
    \epsilon_{1,t}\\
    \vdots\\
    \epsilon_{\bar{n},t}
    \end{pmatrix} , p_t }
    \right \}^{\Bar{t}}_{t=1}
\end{equation}
The standard empirical distribution by definition applies equal probabilities to all historical observations. Applying equal probabilities to historical observations is the equivalent to assuming all historical observations are equally important.

Practitioners tend to rely more on recent observations, and potentially, on additional market information. This lead to Meucci's alternatively specified flexible probabilities which are time and state conditioned. Meucci estimates the future distribution of market invariant non-parametrically \cite{A2013} by using the historical observations of the market invariant as forward looking scenarios, and applying unequal flexible probabilities to these historical scenarios. The flexible probability in this paper is based on how similar the current state of the market is to the state of the market when the historical realization occurred. The state of the market can be measured by any observable state variables such as Consumer Price Index (CPI) and can arise from macro economic, financial, risk and other categories.

Here we set the $\bar{n}$ asset class returns as invariants, and consider a historical time series of $\bar{n}$ joint asset class returns 
$\{{\varepsilon}_{{\bar{n}},{\bar{t}}} = r_{{\bar{n}},{\bar{t}}}\}^{\bar{t}}_{t=1}$.  
Using $m$ state variables we obtain a historical time series of state variables $\{z_{m,t}\}^{\bar{t}}_{t=1}$ and relying on recent observations we derive time and state conditioned flexible probabilities $p_t^{\mathrm{HFP}}$. This flexible probability is then applied to historical asset class returns to estimate the future asset class return distribution:
\begin{equation}\label{eq:2}
    \boldsymbol{r}_{\bar{t}+1} \equiv 
    \begin{pmatrix}
    r_{1,\bar{t}+1}\\
    \vdots\\
    r_{\bar{n},\bar{t}+1}\\
    \end{pmatrix}
   \sim
   \left \{ {\boldsymbol{r}_t \equiv
   \begin{pmatrix}
    r_{1,t}\\
    \vdots\\
    r_{\bar{n},t}\\
    \end{pmatrix}, p_t^{\mathrm{HFP}}
    } \right\}^{\bar{t}}_{t=1}.
\end{equation}
   
In the following sub-sections we briefly summarize some different approaches to obtaining flexible probabilities.

\subsubsection{Time Conditioned Probabilities}

Time conditioned probabilities weight historical returns based on when in time they occurred. We consider two methods of deriving time conditioned probabilities, namely: 1.) rolling window and, 2.) exponentially decayed probabilities.

\subsubsubsection{Rolling window probabilities}

The simplest way to derive flexible probabilities that are time conditioned is the rolling window approach. The method equally weights all historical returns that fall within a rolling window of length $\lambda$ and zero weights all historical returns outside this window, such that:
\begin{equation} \label{eq:3}
    p_t|\bar{t} \equiv p_{t}^{roll} \propto
    \begin{cases}
       1 & \text{if}\,\, t>\bar{t}-\lambda \\
       0 & \text{otherwise}
    \end{cases}. 
\end{equation}
Here the symbol $\propto$ means that the probabilities are re-scaled to sum to 1. A benefit of the rolling window approach is that it easily allows for time-varying estimation. This approach is widely used in the investment industry to give greater importance to recent data than older data. There are key problems with this approach, firstly, it reacts slowly to changes in risk, and secondly, it is quite an abrupt approach where historical returns will either receive full weight or zero weight depending on whether they fall within the targeted window.

\subsubsubsection{Exponentially decayed probabilities}

A smoother approach to derive time conditioned probabilities is to make use of the exponential decay function: 
\begin{equation}\label{eq:4}
p_t|\bar{t} \equiv p_t \propto e^{-\frac{\ln{2}}{\tau}(\bar{t}-t)}.
\end{equation}
Here $\tau$ is the half-life decay of the exponential function which represents the amount of time it would take for the probability to decay and become half the value of the highest value of $\bar{t}$. The higher the half-life, the higher the weight to recent data. In contrast to rolling window approach, the exponential decaying approach is more responsive to changing risk and offers a smoother profile where historical returns are given a higher weight depending on how recent it is. Time conditioning based on exponential decayed probabilities was first introduced by Bourdoukh et al \cite{A1998}.

\FloatBarrier
\begin{figure}[h]
\centering
\includegraphics[width=0.5\textwidth]{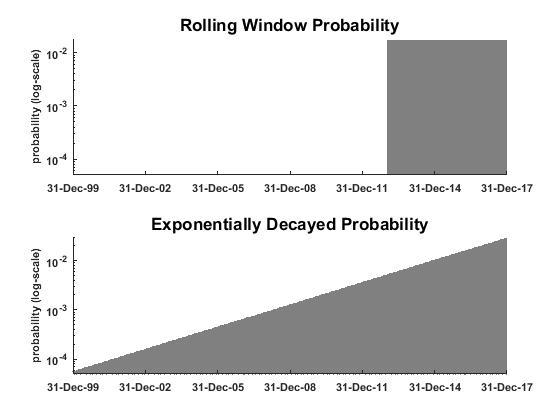}
\caption{Shows an example of how to obtain time conditioned probabilities using: 1.) the rolling window approach where $\lambda$ = 60 months, and  2.) an exponential decay approach where $\tau$= 24 months. The grey region represents the weight or probability applied relative to the corresponding calendar date.}
\label{fig:fig1}
\end{figure}
\FloatBarrier

\subsubsection {State Conditioned Probabilities}

\begin{figure}[h]
\centering
\includegraphics[width=0.5\textwidth]{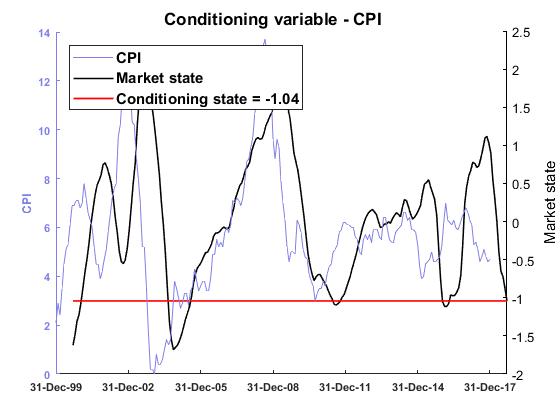}
\caption{Here we consider month-on-month changes in inflation, CPI (Table \ref{table:Table3}), as a state variable. We plot the raw CPI on $y_1$ axis which we then smooth, score and finally standardize into the market state variable which we plot on $y_2$ axis, and condition by target state CPI=-1.04, which is the latest CPI market state value as shown by the red line.}
\label{fig:fig3}
\end{figure}

State conditioned probabilities give a greater weight to historical returns where market conditions display particular desired characteristics. The approach can be used to give a greater weight to history that is most similar to market conditions prevailing today or for stress testing purposes by giving a greater weight to history that is most similar to specific stressed market conditions. In Figure \ref{fig:fig3} we show how CPI can be used as a state variable and use the latest CPI state value as the desired market condition we want to stress. We consider two methods of deriving state conditioned probabilities, namely: 1.) crisp and, 2.) kernel conditioned probabilities.

\subsubsubsection {Crisp conditioned probabilities}

The simplest way to state condition is by using crisp conditioned probabilities. In this approach historical returns are given an equal weight if the state variable historically falls within a specified $\alpha$\ range of the target state variable $z^{*}$ and zero weight to all other returns. The crisp probabilities can be written as:
\begin{equation}\label{eq:5}
    p_t|z^{*} \equiv p_{t}^{crisp} \propto
    \begin{cases}
       1 & \text{if}\,\,z_t \in \mathcal{R}(z^{*})\\
       0 & \text{otherwise}
    \end{cases}.
 \end{equation}

Similar to the rolling window approach the crisp conditioning is an abrupt approach where historical returns will either receive full weight or zero weight depending on whether they fall within or outside the range of the targeted state.

\subsubsubsection {Kernel conditioned probabilities}

The smoother method to time condition is Kernel conditioning. The approach is to weight historical returns based on the distance between the state variable at the point in time and the target state variable. The distance is measured using either an exponential or a Gaussian kernel; we can then write the probabilities as:
\begin{equation}\label{eq:6}
p_t|z^{*} \equiv p_{t}^{\mathrm{ker}} \propto e^{\frac{-|z_t-z^{*}|^{\gamma}}{h}}. 
\end{equation}
Here the constant $h$ is the bandwidth of the kernel and determines the smoothness of the kernel and hence the smoothness of the probabilities.  The constant $\gamma$ indicates the type of kernel chosen, where $\gamma = 1$ corresponds to an exponential kernel and $\gamma = 2$ corresponds to a Gaussian kernel. 

\FloatBarrier
\begin{figure}[h]
\centering
\includegraphics[width=0.5\textwidth]{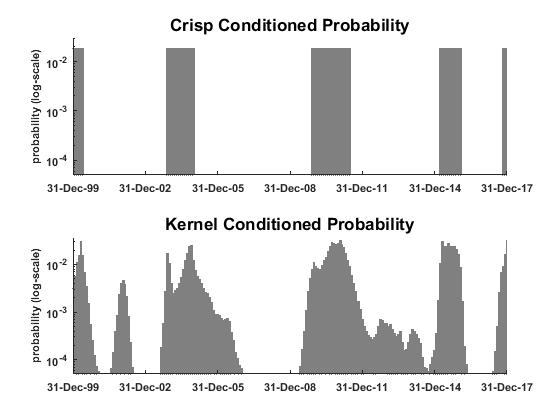}
\caption{Shows an example of how to obtain state conditioned probabilities using: 1.) crisp conditioning probability, and 2.) kernel conditioning probability with h = 0.5 and $\gamma$ = 2. The state variable used to condition is CPI and the target state is CPI=-1.04. The grey region represents the weight or probability applied relative to the corresponding calendar date.}
\label{fig:fig2}
\end{figure}
\FloatBarrier

\subsubsection {Time and State Conditioned Probabilities}

As mentioned in the previous subsections we can derive flexible probabilities that are conditioned by time or conditioned by state. But, both time and state conditioning are important, so it would be useful to derive flexible probabilities that are conditioned by both time and state. By doing so we can give higher weight to historical returns that are more recent and have similar characteristics to desired market conditions.

Meucci \cite{A2013} demonstrates that the HS-FP approach \cite{B2008} can be used to derive time and state conditioned flexible probabilities by utilizing some of the aforementioned techniques. This approach delivers a forward looking or posterior distribution that reflects most recent and desired market conditions with least distortion to prior distribution. However, we need to be able to quantify how different one distribution is from another. 

\subsubsubsection {Entropy-based conditioning procedure}

It is through relative entropy criterion that we can quantify how different one distribution is from another, and by minimizing the relative entropy we obtain a forward looking distribution that is as close as possible to the prior distribution. Meucci \cite{A2013} applies the following steps to ensure that the distribution is both time and state conditioned.

First, he specifies a state variable z with a target state value for this state variable $z^{*}$ and sets the initial flexible probabilities as the crisp state conditioned probabilities:
\begin{equation}\label{eq:7}
    p_t|z^{*} \equiv p_{t}^{\mathrm{crisp}} \propto
    \begin{cases}
       1 & \text{if}\,\,z_t \in \mathcal{R}(z^{*})\\
       0 & \text{otherwise}
    \end{cases}. 
 \end{equation} 
 
Second, he calculates the "crisp" mean and standard deviation of the target state variable as below:
\begin{equation}\label{eq:8}
\mu|z^{*} = \sum_{t = 1}^{\bar{t}}z_{t}p_{t}^{\mathrm{crisp}} \mbox{ and } \sigma|z^{*} = \sqrt{\sum_{t = 1}^{\bar{t}}z_{t}^{2}p_{t}^{\mathrm{crisp}}- (\mu|z^{*})^{2}}
\end{equation}

Third, the views are expressed. The views are statements on target state variable that distort the prior distribution in the least spurious way. 
The view is that the yet-to-be determined flexible probabilities $p_t$ must match the moments of the crisp state conditioned probability:
\begin{equation}\label{eq:9}
    \nu|z^{*} \equiv 
    \begin{cases}
       \sum_{t = 1}^{\bar{t}}z_{t}p_{t} = \mu|z^{*} &  \\
       \sum_{t = 1}^{\bar{t}}z_{t}^{2}p_{t} \leq (\mu|z^{*})^{2}+(\sigma|z^{*})^{2}.& 
    \end{cases} 
 \end{equation}
 
Finally, he sets prior distribution as the exponentially decayed time conditioned probabilities
\begin{equation}\label{eq:10}
p_t|\bar{t} \equiv p_t \propto e^{-\frac{\ln{2}}{\tau}(\bar{t}-t)}
\end{equation}
and calculates the posterior distribution that satisfies views by minimizing the relative entropy or Kullback-Leibler divergence \cite{B1951} between the exponential decayed time conditioned probability prior in \ref{eq:10} and flexible probability distributions while satisfying crisp state moment views in (\ref{eq:9})
\begin{equation}\label{eq:11}
\textbf{p}^|z^{*} \equiv \argmin_{\textbf{p} \in \nu|z^{*}} \epsilon(\textbf{p},\textbf{p}^{exp})
\end{equation}
where
\begin{equation}\label{eq:12}
\epsilon(\textbf{p},\textbf{p}^{exp}) = \sum_{t=1}^{\bar{t}}p_t\ln\bigg(\frac{p_t}{p_{t}^{exp}}\bigg).
\end{equation}
The final flexible probability as explained in Meucci \cite{A2013} is a combination of the exponential decay prior (\ref{eq:4}) and kernel conditioning with optimal bandwidth and center. As a result of setting an inequality view on the second moment in (\ref{eq:9}) the kernel estimator is allowed to switch freely between the best exponential kernel and best Gaussian kernel.

\FloatBarrier
\begin{figure}[h]
\centering
\includegraphics[width=0.5\textwidth]{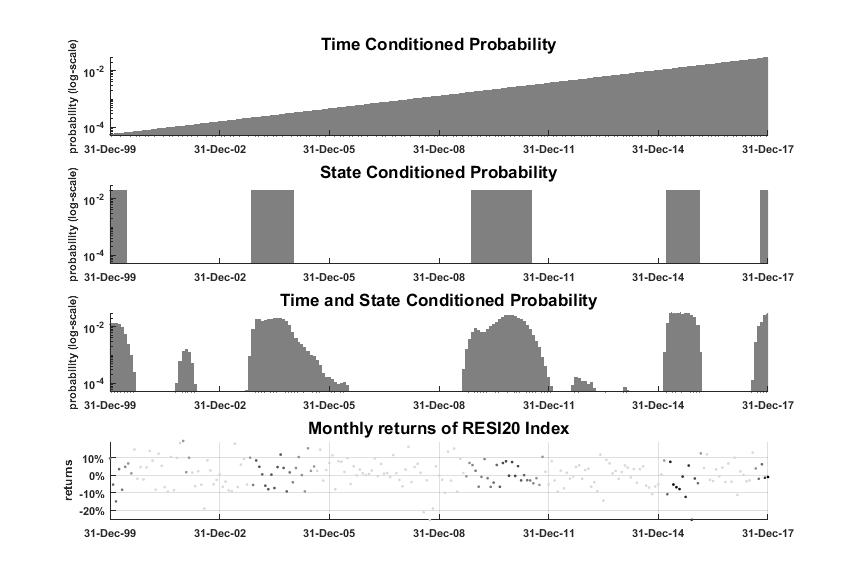}
\caption{The first three panels show an example of obtaining time, state, time and state conditioned probabilities using a single state variable. The state variable used to condition is CPI and the target state is CPI=-1.04. The grey region represents the weight used relative to the corresponding calendar date.The first panel shows time conditioning by exponential decayed probability where $\tau$= 24 months, the second panel shows state conditioning by crisp conditioning, the third panel shows time and state conditioning using entropy pooling approach. Finally the last panel shows how the Resource 20 Index returns (see Table \ref{table:Table1}) can be conditioned by time and state using entropy pooling approach.}
\label{fig:fig4}
\end{figure}
\FloatBarrier

\subsection{Combining multiple state variables} \label{sec:fp-mult}

In practice, there is greater benefit in conditioning by multiple state variables \cite{B1969} than by a single state variable. However, in order to do so we need to calculate the time and state conditioned probability for each state variable and decide how much to weight each probability: 
\begin{equation}\label{eq:13}
\textbf{p}_{comb} = w_{1}\textbf{p}|z_{1}^{*}+\ldots+w_{\bar{q}}\textbf{p}|z_{\bar{q}}^{*}.
\end{equation}
In this paper we consider two combination methodologies equal weighting and DCC weighting.

\subsubsection {Equal weighting}

The simplest approach to weighting the probabilities of the different state variables is equal weighting: 
$EQ_q = \frac{1}{q},$ where $q = 1,\ldots,\bar{q}$ is the number of state variables.

The equal weighting approach is straight forward and has been argued by many as a hard combination method to beat \cite{C2006} and \cite{C2010}.\\

\subsubsection {DCC weighting}

In this approach the flexible probability from each state variable is weighted based on the Degree of Conditioning and Correlation (DCC) between the different state variables \cite{A2012}. If a given state variables is lowly correlated with the other state variables, or imposes a smaller degree of conditioning on many historical scenarios, then this state variable will receive a higher weighting.

To measure the degree of conditioning from a given state variable the Effective Number of Scenarios introduced by Meucci in \cite{A2012} is used: 
\begin{equation}\label{eq:14}
\mathcal{T} = e^{-\sum_{t = 1}^{\bar{t}}p_t\ln{p_t}}.
\end{equation}
When all historical scenarios are equally weighted, $p_t = \frac{1}{\bar{t}}$, then the degree of conditioning is minimal and the Effective Number of Scenarios is maximal at $T = \bar{t}.$ On the other hand, if only one historical scenario is assigned all probability, then the degree of conditioning is maximal, and the Effective Number of Scenarios is minimal with $T = 1$.

To measure the degree the correlation between a given state variable and the rest of the state variables the following steps are followed. Firstly the Bhattacharyya coefficient \cite{C1943} is used to measure the degree of correlation between the different state variables  for any pair of Flexible probabilities $(p_q, p_r)$ as
\begin{equation}\label{eq:15}
b_{q,r} \equiv \sum_{t = 1}^{\bar{t}}\sqrt{p_{t,q}p_{t,r}}
\end{equation} 
then the Hellinger distance \cite{C1909} is calculated as
\begin{equation}\label{eq:16}
d_{q,r} \equiv \sqrt{1-b_{q,r}}.
\end{equation}
Finally, the diversity index \cite{C1958}, $\mathcal{D}_{q}$ is used to summarise the degree of similarity between the probabilities from a given state variable and the probabilities from the others state variables. The diversity index is basically the average of Hellinger distances between the given set of probabilities and the remaining probabilities:
\begin{equation}\label{eq:17}
\mathcal{D}_{q} = \frac{1}{\bar{q}-1}\sum_{r \neq q}^{}d_{q,r}.
\end{equation}

The final weighting for the flexible probability from each state variable is then calculated as 
\begin{equation}\label{eq:18}
\mathrm{DCC}_q = \frac{\mathcal{T}_q\mathcal{D}_q}{\sum_{r=1}^{\bar{q}}\mathcal{T}_r\mathcal{D}_r}
\end{equation}
where $q = 1,\ldots,\bar{q}$. The ${p}_{comb}$ that is derived weighting each state variable by DCC is referred to as the Ensemble Flexible Probability ${p}_{ensemble}$.

\FloatBarrier
\begin{figure}[h]
\centering
\includegraphics[width=0.5\textwidth]{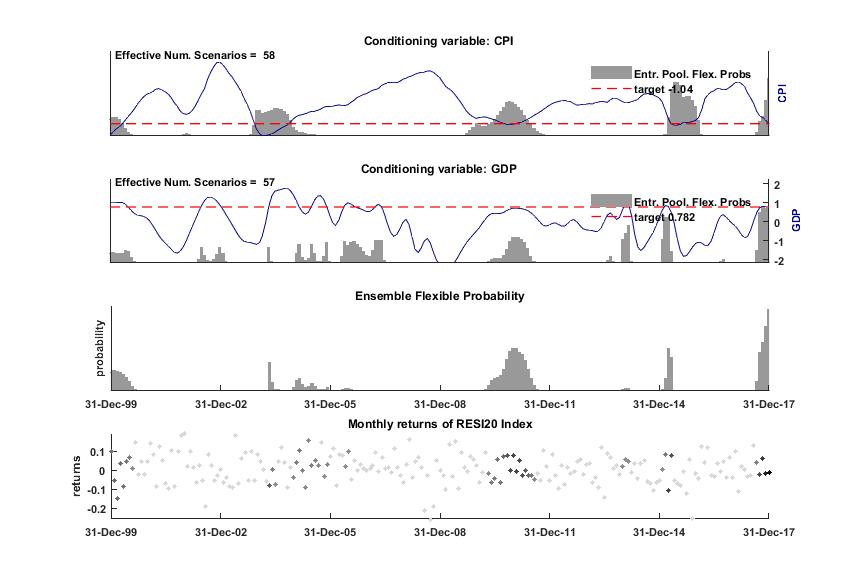}
\caption{Shows an example of how to obtain time and state conditioned probabilities using multiple state variables CPI and GDP. In the first panel we condition only by the latest market state of CPI state variable and overlay the time and state conditioned probability. In the second panel we condition only by the latest market state of GDP state variable (see Table \ref{table:Table3}) and overlay the time and state conditioned probability. In the third panel we show the ensemble probability when conditioning by both time and multiple state variables. Finally in the last panel we show the Resources 20 Index (see Table \ref{table:Table1}) conditioned by both time and the multiple state variables CPI and GDP.}
\label{fig:fig5}
\end{figure}
\FloatBarrier

\section{Benchmark Models} \label{sec:bench}

The following benchmark models will be considered in the empirical study.

\subsection{Equal Weighted (EW)}

Equal weighted or 1/n portfolios are very straight forward and have been widely used by investment practitioners \cite{BM2001,BM2004}. This allocation strategy has also been cited in numerous academic literature to be efficient out-of-sample \cite{BM2009} and hard to beat \cite{C2006} and \cite{C2010}, supporting the usage of EW as a benchmark model in the paper. 

\subsection{Classic Mean Variance Optimization (MVO)}

In Markowitz’s \cite{A1952} Nobel winning Mean Variance portfolio optimization theory the mean and covariance of asset class returns are assumed to be known. In reality, however these moments are unknown and commonly estimated from historical data. In this paper the classic Mean Variance benchmark model assumes the predicted mean of the next period’s return distribution is taken as the current historical mean return, this is the equivalent to the HS-FP model assigning an equally weighted probability to historical returns. Many academic papers \cite{A2008,BM1999,BM2003} have argued that most of the traditional return predictors fail to outperform the historical average, supporting the usage of classic MVO portfolios as another benchmark model in the paper. 

\section{Implementation} \label{sec:implement}

\subsection{Data}

We use monthly data of five asset classes and ten state variables for the past approximately 19 years {\it i.e.} Feb 1998 to Dec 2017. 

\subsubsection{Data on asset classes}

In this section the asset classes considered in this paper are introduced. Table \ref{table:Table1} below lists the names of the asset classes, proxies and codes used to extract from Bloomberg and Table \ref{table:Table2} shows the return, risk and Sharpe ratio of asset classes where the risk free rate is $7.25\%$ over the period Feb 1998 to Dec 2017. 

\begin{table}[H]
\begin{center}
\begin{tabular}{|p{2cm}|p{2cm}|p{4.5cm}|}
\hline
 Asset Class & Code & Proxy\\
 \hline
 Resource & RESI20 Index & FTSE/JSE Africa Resource 10 Index \\
 Industrial & INDI25 Index & FTSE/JSE Africa Industrial 25 Index \\
 Financial & FINI15 Index & FTSE/JSE Africa Financial 15 Index \\
 All Bond  & ALBTR Index & ALBI Total Return Index \\
 USDZAR Currency & USDZAR Curncy & United State Dollar/South African Rand Cross \\
 \hline
\end{tabular}
\caption{Asset Classes: The list of asset classes is given with their Bloomberg codes and the proxies used to represent these asset classes.}
\end{center}
\label{table:Table1}  
\end{table}

\begin{table}[H]
\begin{center}
\begin{tabular}{|p{4cm}|p{1.5cm}|p{1.5cm}|p{1.5cm}|}
\hline
Asset Class & Ann. Return & Ann. Volatility & Sharpe Ratio \\
 \hline 
Resources  & 11.16\% & 12.12\% & 0.32 \\
Industrial  & 8.27\% & 6.26\% & 0.16 \\
Financial  & 8.13\% & 10.91\% & 0.08 \\
Bonds  & 8.27\% & 6.26\% & 0.16 \\
Currency  & 8.13\% & 10.91\% & 0.08 \\
\hline
\end{tabular}
\caption{Indicative Asset Class Performance: The annualised returns, annualised volatilities, and finally, annualised Sharpe ratios are provided as summary statistics for the asset classes over the period February 1998 until December 2017.}
\end{center}
\label{table:Table2}  
\end{table}

\subsubsection{Data on state variables}

In this section we list the state signals, these state signals are processed to form state variables. We use 10 state variables, these range from macroeconomic, risk and trend based categories. These variables were largely selected based on : the existence of academic literature supporting the usage of these variables, if these variables where commonly used by investment practitioners and if historical data was readily available for each of these state variables for the specified analysis period.

\begin{table}[H]
\begin{center}
\begin{tabular}{|p{5.0cm}|p{0.5cm}|p{3.0cm}|}
\hline
 State Signal & Freq. & Code(s) used \\
 \hline
 SA Real Gross Domestic Product & Q. & SAGDPANN Index \\
SA Domestic Leading Indicator & M. & OEZAKLAP Index\\
SA Consumer Price Index & M. & SACPIYOY Index\\
SA Money Supply & M. & SAMYM3Y Index\\
SA Equity Momentum & M. & JALSH Index \\
JP Morgan EMBI Index & M. & JPEIGLBL Index \\
USD ZAR Currency & M. & USDZAR Curncy\\
USA PMI & M. & NAPMPMI Index\\
USA VIX Index & M. & VIX Index \\
Equity Risk Premium \ (USA-SA) & M. & SPX Index, US0003M Index, JALSH Index, JIBA3M Index \\
\hline
\end{tabular}
\caption{State Variable Signals: The list of state variables used as investment signals, along with their reporting frequency and Bloomberg codes are provided; these include variables from both domestic and global econometric candidate signals and cover macroeconomic, risk and trend based variables. The list is indicative and not exhaustive. The sampling frequencies are Monthly (M) and quarterly (Q).}
\end{center}
\label{table:Table3}  
\end{table}

All state signal data was sourced from Bloomberg. The Bloomberg codes listed above where used to pull data for state signal 1-4 and 6-9. For state signal 5, the 12 month rolling return of the FTSE JSE All share Index was used as a proxy for SA Equity Momentum. For state signal 10, the differential between the (S$\&$P 500 less 3 month LIBOR) and (FTSE JSE All Share Index less 3 month JIBAR) was used as a proxy of equity risk premium differential. All state signals except for GDP have monthly frequency hence cubic-spline interpolator was used to convert quarterly GDP data into monthly data. In order to account for the delay in receiving data we lag: GDP data by 3 months, the OECD Domestic Leading Indicator by 6 months, CPI by 1 month and Money Supply by a month.

In academic literature such as \cite{E2013} and \cite{A2011} GDP and Inflation are commonly used together to understand the economic state of a country. Investment practitioners Munro and Silberman \cite{Z2008} use these two variables together to classify the current regime of the South African economy. The SARB uses the long run trend of GDP to identify the business cycle \cite{Z2005}, and \cite{Z2011} examines the effect of macroeconomic variables including GDP on the South African equity market and found South Africa’s stock market index is positively influenced by the growth rate of real GDP.

The paper also considers the South African OECD Domestic Leading Indicator, this indicator is widely used by practitioners and aims to give a sense of future GDP growth by approximately 6 months, \cite{Z2015} shows that Composite Leading Indicators has significant in-sample relationship in describing the course of the economy. 

Inflation has been cited by both local and global research to have asset class return predictability. Gupta and Modise \cite{Z2111} examined the effect of macroeconomic variables including Inflation on the South African equity market and found that the inflation rate shows a strong out-of-sample predictive power from 6-months-ahead horizons. In the U.S. \cite{Z2006} shows that inflation has substantial out-of-sample forecasting abilities for real US stock returns. 

Money Supply has been shown to display out-of-sample equity return predictability \cite{Z2012}, locally \cite{Z2111} shows money supply has in-sample predictive power in the short run.

Jostova \cite{Z2003} showed that emerging market spreads could be used to predict emerging sovereign debt markets returns . 

The relationship between stock returns and exchange rates was analyzed in \cite{Z2000} and showed a relationship exist between the stock returns and exchange rates due to real interest rate disturbance and inflationary disturbance.

There is significant predictability from past returns \cite{Z2014} and \cite{Z2093} also document that momentum investments earned abnormal returns. 

The US ISM Manufacturing PMI Composite Index is a survey-based indicator that is widely followed by investors. This indicator has been investigated by many \cite{Z1991,Z1993,Z2114} and could have some return predictability.

In \cite{Z2018} the authors show that VIX plays a role in the relationship between idiosyncratic volatility and stock returns such that when VIX decreases there is a decrease in risk aversion resulting in improved stock returns. 

In \cite{Z2112} the authors show that lagged U.S. returns significantly predict returns in numerous non-U.S. countries substantiating the usage of equity risk premium differential relative to the US and showcasing the leading role that United States plays in return predictability of other countries.

It should be noted that the paper uses a couple of state variables sourced from the US or relative to the US i.e. the US PMI, US VIX, Equity Risk Premium (ERP) differential and Currency relative to the US. The US is the major driver of equilibrium relations in eleven emerging Asian-Pacific stock markets \cite{Z2002} which suggests the leading role that United States plays in other emerging market countries and potentially South Africa.

\subsection{Methodology}

In this section we detail the steps required to convert the data in the previous subsection into the required format for the HS-FP model. We discuss how the HS-FP Model is used to forecast asset class return and risk, how we select the optimal portfolio and finally how the backtesting is setup to calculate the out-of-sample returns.

\subsubsection{Data preparation}

First, we need to determine the historical time series of invariants $e_t$. The asset class index levels in Table \ref{table:Table1} are retrieved and the logarithmic return series $r_t$ is calculated. The historical time series of invariants $e_t$ is estimated by $r_t$  as seen below: 
\begin{equation}\label{eq:19}
\{{\varepsilon}_{{\bar{n}},{\bar{t}}} = r_{{\bar{n}},{\bar{t}}} = \ln(v_{{\bar{n}},{\bar{t}}}/v_{{\bar{n}},{\bar{t}}-1})\}^{\bar{t}}_{t=1}.
\end{equation}
Second, we need to determine the historical time series of state variables $z_{m,t}$. In Table \ref{table:Table3} we provide a list of m state signals $\{S_{m,t} \equiv (S_{1,t},\hdots, S_{m,t}) \}^{t=\bar{t}}_{t=1}$ used in this paper. State signals are typically noisy so we need to smooth and score them to obtain a historical time series of state variable  $\{z_{m,t} \equiv (z_{1,t},\hdots, z_{m,t})\}^{t=\bar{t}}_{t=1}$ that we can compare and combine easily. 

Starting with signal time series $S_{m,t}$ we first smooth this series to obtain $z_{m,t}^{\mathrm{smooth}}$ by calculating the exponential weighted moving average where we set the fast half life and slow half life parameters at 3 and 12 months respectively in Section \ref{sec:analysis}.
\begin{equation}\label{eq:20}
z_{m,t}^{\mathrm{smooth}} = \mathrm{smooth}(S_{m,t}) \equiv \mathrm{EWMA}_{w}^{\tau_{HL}}(S_{m,t}).
\end{equation}
Then in order to compare how this smoothed signal has evolved over time, we standardize by calculating the z-score of $z_{m,t}^{\mathrm{smooth}}$. This then forms the state variable time series $z_{m,t}$ required in HS-FP model.
\begin{equation}\label{eq:21}
z_{m,t} = \mathrm{score}(z_{m,t}^{\mathrm{smooth}}) \equiv \frac{z_{m,t}^{\mathrm{smooth}}-{\mu}(z_{m,t}^{\mathrm{smooth}})}{{\sigma}(z_{m,t}^{smooth})}.
\end{equation}

In Section \ref{sec:analysis} of the paper we use monthly data spanning approximately 19 years from which we use 5 years as an initial training window which grows with the addition of new information each subsequent month.

The target state variable $z_t^{*}$ at each update of the model is set as the most recent value of the state variable. The leeway parameter which is the +/- range from the current value of the state variable is set at 0.1 throughout Section \ref{sec:analysis}. 

\subsubsection{Forecasting return and covariance}

We show how to use the HS-FP model to forecast the expected return $\mathbb{E}[R]$ and covariance $\mathbb{C}\mathrm{ov}[R]$ of asset classes listed in Table \ref{table:Table1}.

The HS-FP model assigns probabilities to historical returns over the growing window each month conditioned by time and either single or multiple state variables. This flexible probability adjusted return and covariance then forms each asset class predicted mean and covariance for the next period’s return distribution.

At time t having obtained $r_t$, $z_t$ and specified the target state variable as $z_t^{*}$, we then use the HS-FP model detailed in Section \ref{sec:hs-fp} to obtain the time and state conditioned flexible probabilities $p_t^{\mathrm{HFP}}$ for each asset class. 

Where the estimated $\mathbb{E}[R]$ and covariance $\mathbb{C}\mathrm{ov}[R]$ for each asset class is given by:  
%\begin{equation*}
%    \{\Gamma_t,\mathrm{p}^{\mathrm{HFP}}_{t}\}^{\Bar{t}}_{t=1}
%\end{equation*}
\begin{eqnarray}\label{eq:22}
    \mathbb{E}[R]&=&\sum_{t=1}^{\Bar{t}}\mathrm{p}^{\mathrm{HFP}}_{t}\mathrm{r}_{\bar{t}}, ~~\text{and} \\
    \mathbb{C}\mathrm{ov}[R]&=&\sum_{t=1}^{\Bar{t}}\mathrm{p}^{\mathrm{HFP}}_{t}\mathrm{r}_{t}\mathrm{r}_{t}'-\left[\sum_{t=1}^{\Bar{t}}\mathrm{p}^{\mathrm{HFP}}_{t}\mathrm{r}_{t}\right]\left[\sum_{t=1}^{\Bar{t}}\mathrm{p}^{\mathrm{HFP}}_{t}\mathrm{r}_{t}\right]'. \nonumber
\end{eqnarray}

\subsubsection{Portfolio optimization and backtesting}

Having obtained the forward looking distribution for the next period’s asset class returns we then need to select the optimal asset allocation. We use the Markowitz framework \cite{A1952} and select the maximum Sharpe ratio \cite{E1994} as the optimal portfolio. The only constraints used in the portfolio optimization step is that the portfolio weight need to be non-negative and to sum to 1 (long only constraint). No upper or lower bounds or group bounds where specified for asset classes weights and no turnover and tracking error constraints where specified.

In this paper we use monthly data spanning approximately 19 years from which we use 5 years i.e. Feb 1998 to Feb 2003 as an initial training window which grows with the addition of new information each subsequent month and hence the out-of-sample returns span from March 2003 to Dec 2017 which is about 14 years. Estimates are updated monthly upon the arrival of new data. The forecast horizon is fixed at one month and the rebalancing frequency is half yearly. 

The backtesting process is as follows, the initial training window is used to estimate the first $\mathbb{E}[R]$ and $\mathbb{C}\mathrm{ov}[R]$ for each asset class, then the efficient frontier is plotted and we select the optimal portfolio as the portfolio from the efficient frontier with maximum Sharpe ratio, the optimal portfolio is fixed and tested on the out-of-sample returns and after 6 months the process repeats.

\section{Results} \label{sec:analysis}

In order to understand the results we report on a number of summary statistics, observe the optimal portfolio weights over time, and cumulative and relative rolling return graphs. The statistics we report are popular in the investment industry and consist of the annualised gross returns, annualised volatility, Sharpe ratios, maximum draw down, average monthly turnover and conditional value at risk.

In the first subsection we compare the out-of-sample results when conditioning by each of the 10 state variables individually. In the second subsection  we then combine all the state variable and compare the results of the different combination methodologies covered in this paper the equal weighting (EQ) and weighting by Degree of Conditioning and Correlation (DCC) and select the best performing amongst these as the HS-FP model going forward. Finally in the last subsection, the performance of the HS-FP model is compared to benchmark models namely the classic MVO and EW. The risk free rate is $7.25\%$ over the analysis period.

The analysis of the results is based on an out-of-sample performance, meaning we evaluate the performance of the HS-FP and benchmark models on a period of data which is different from the period of data we used to identify the optimal asset allocation. The out-of-sample returns in Section \ref{sec:analysis} assumes zero transaction costs and make no adjustment for indirect costs related to portfolio rebalancing. 

\subsection{Comparing results of single state variables}\label{sec:single-var}

In this subsection we report the results where each of the state variables are used individually in forecasting all asset classes next period return distribution. 

Looking at these performance, risk and turnover statistics in Table \ref{table:Table4}. When annualised gross returns are observed the best performers are SA Inflation, US PMI and ERP differential. In terms of the risk statistics, as seen in Table \ref{table:Table5}: US VIX, EMBI and ERP differential have the the lowest annualised volatility, whilst VIX, USD/ZAR exchange rate and ERP differential have the lowest Maximum Drawdown and VIX, ERP differential and US PMI have the lowest CVaR. Also VIX, Momentum and ERP differential have the lowest average monthly turnover. Finally in terms of Sharpe ratios, as seen in Table \ref{table:Table4}: US PMI, ERP differential and SA Inflation are the state variables that have the highest Sharpe ratios. 

Different state variables do well in the different categories, there is no single state variable that ranks best across all categories. That being said, US VIX and ERP differential perform well across all the different risk measures and ERP differential is the only state variable that ranks in the top 3 position across all categories.

\FloatBarrier
\begin{table}                                                           
\centering                                                              
\begin{tabular}{|p{4cm}|p{1.5cm}|p{1.5cm}|p{1.5cm}|}                                              
\hline                                                                  
State Variables & Ann. Return  & Ann. Volatility & Sharpe Ratio \\                                
\hline                                                                  
SA Inflation & 9.19 & 10.25 & 0.18 \\            
\hline                                                                  
SA GDP & 6.03 & 8.39 & -0.14 \\                  
\hline                                                                  
Domestic Lead Indicator & 8.33 & 8.54 & 0.12 \\ 
\hline                                                                  
EMBI & 7.03 & 6.84 & -0.03 \\            
\hline                                                                  
USDZAR Currency & 8.58 & 7.47 & 0.17 \\         
\hline                                                                  
ERP differential & 8.87 & 6.88 & 0.23 \\                     
\hline                                                                  
US PMI & 9.03 & 7.05 & 0.25 \\                  
\hline                                                                  
VIX & 7.96 & 6.66 & 0.10 \\                    
\hline                                                                  
Money Supply & 7.92 & 10.05 & 0.06 \\            
\hline                                                                  
Momentum & 5.73 & 9.37 & -0.16 \\                
\hline                                                              
\end{tabular}                                                           
\caption{Single State Variable Summary Statistics I: The state variables listed are used individually as the only state variable to condition the HS-FP model; popular summary statistics such as gross annualised return, annualised volatility and annualised Sharpe ratio are shown for the March 2003 - Dec 2017 out of sample period.}                                               
\label{table:Table4}                                                
\end{table} 

\begin{table}                                                           
\centering                                                              
\begin{tabular}{|p{4cm}|p{1.5cm}|p{1.5cm}|p{1.5cm}|}                                              
\hline                                                                  
State Variables & Max. Drawdown  & CVaR & Turnover \\                                
\hline                                                                  
SA Inflation & 23.78 & 7.64 & 6.64 \\            
\hline                                                                  
SA GDP & 28.48 & 8.17 & 4.63 \\                  
\hline                                                                  
Domestic Lead Indicator & 15.06 & 6.73 & 6.30 \\ 
\hline                                                                  
EMBI & 10.54 & 4.75 & 6.42 \\            
\hline                                                                  
USDZAR Currency & 9.53 & 4.55 & 5.32 \\         
\hline                                                                  
ERP differential & 10.07 & 4.01 & 5.01 \\                     
\hline                                                                  
US PMI & 12.53 & 4.04 & 5.86 \\                  
\hline                                                                  
VIX & 7.44 & 3.47 & 4.07 \\                    
\hline                                                                  
Money Supply & 28.48 & 8.21 & 5.43 \\            
\hline                                                                  
Momentum & 29.46 & 9.21 & 4.88 \\                
\hline                                                              
\end{tabular}                                                        
\caption{Single State Variable Summary Statistics II: The state variables listed are used individually as the only state variable to condition the HS-FP model; summary statistics such as maximum drawdown, conditional value at risk and average monthly turnover are calculated. These statistics are calculated as they look into historical rare events and assess the implication for turnover from conditioning by a single state variables. These summary statistics are shown for the March 2003 - Dec 2017 out of sample period.} 
\FloatBarrier
                                             
\label{table:Table5}                                                
\end{table} 

\subsection{Comparing results of multiple state variable} \label{sec:multi-var}

In this subsection we report the results where all state variable mentioned in the previous sections are used to forecast all asset classes next period return distribution. We allocate to these multiple state variables based on two combination methods EQ and DCC.

\FloatBarrier
\begin{table}[H]
\begin{center}
\begin{tabular}{|p{4cm}|p{1.5cm}|p{1.5cm}|p{1.5cm}|}
\hline
 Combination method & Ann. Return & Ann. Volatility & Sharpe Ratio \\
 \hline
 EQ   & 8.26\% & 6.08\% & 0.16 \\
 DCC  & 11.16\% & 12.12\% & 0.32 \\
 \hline
\end{tabular}
\caption{Multiple State Variable Summary Statistics I: The state variables listed in Table \ref{table:Table3} are used in combination to condition the HS-FP model; two methods to combine information from each state variable is assessed; equal weighting (EQ) where the conditioned probability from each state variable is equally weighted to give final probability and weighting the conditioned probability from each state variable by Degree of Conditioning and Correlation (DCC) between the different state variables. Popular summary statistics such as gross annualised return, annualised volatility and annualised Sharpe ratio are shown for the March 2003 - Dec 2017 out of sample period.}
\end{center}
\label{table:Table6}  
\end{table}

\begin{table}[H]
\begin{center}
\begin{tabular}{|p{4cm}|p{1.5cm}|p{1.5cm}|p{1.5cm}|}
\hline
 Combination method & Max. Drawdown & CVaR & Turnover\\
 \hline
 EQ   & 5.80\% & 3.58\% & 2.60 \\
 DCC  & 21.20\% & 9.84\% & 10.93 \\
 \hline
\end{tabular}
\caption{Multiple State Variable Summary Statistics II: The state variables listed in Table \ref{table:Table3} are used in combination to condition the HS-FP model; two methods to combine information from each state variable is assessed; equal weighting (EQ) where the conditioned probability from each state variable is equally weighted to give final probability, and weighting the conditioned probability from each state variable by Degree of Conditioning and Correlation (DCC) between the different state variables. Summary statistics such as maximum drawdown, conditional value at risk and average monthly turnover are calculated. These statistics are calculated as they look into historical rare events and assess implication for turn-over from conditioning by multiple state variables. These summary statistics are shown for the March 2003 - Dec 2017 out of sample period.}
\end{center}
\label{table:Table7}  
\end{table}
\FloatBarrier

Firstly, we note that based on the results in Table \ref{table:Table6} and Table \ref{table:Table7} combining multiple state variables improves return, risk and turnover statistics in comparison to single state variable approach as seen in Table \ref{table:Table4} and Table \ref{table:Table5}. For example, the SA Inflation state variable records the highest annualised gross return $9.19\%$ from the singles state variable but the DCC approach of combining multiple state variables records $11.16\%$. Looking at risk and turnover in Table \ref{table:Table7} the EQ approach has lower volatility and turnover than any single state variable. Finally, when risk adjusted returns are observed the Sharpe ratio of DCC is higher than any single state variable. These result highlight the benefit of using multiple state variables for improved diversification and returns.

Now, when comparing the combination methodologies, the best performing combination method in terms of annualised gross returns is weighting by DCC. However weighting by EQ has better risk and turnover statistics. Finally in terms of risk adjusted returns the Sharpe ratio is the highest for the combination method DCC. Due to this we refer to time and conditioning by multiple state variables where the weighting to multiple state variables is derived by DCC as the HS-FP going forward.

\subsection{Comparing out of sample HS-FP results to benchmark models}

In this section we compare results of HS-FP to benchmark models. We compare allocations over time, summary performance, risk and turnover statistics, as well as rolling return profiles to observe if there are consistent differences between HS-FP approach and the benchmark models.

\subsubsection{Comparison of allocations over time}

Comparing the asset allocations of HS-FP to classic MVO and EW benchmark models over time the following insights are observed.

The HS-FP and naturally EW approach does allocate at least once to all the asset classes whilst the classic MVO never allocates to financials and currency. The least favorite asset classes on average for the HS-FP approach is also financials and currency. Hence both HS-FP and classic MVO are underweight financials and currency relative to EW approach.

The fixed income position is the largest allocation for both HS-FP and classic MVO on average over time. The classic MVO has a considerably larger allocation to fixed income on average than HS-FP. Hence HS-FP and classic MVO is overweight fixed income relative to EW approach. 

During the Global Financial Crisis (2007-2008) the HS-FP approach had a greater overweight to fixed income and industrials and greater underweight to financials and currency relative to EW approach. Meanwhile the HS-FP approach was overweight industrials and underweight fixed income (despite this asset class being HS-FP biggest average allocation during this period) relative to classic MVO. 

Currently as at Dec 2017, the HS-FP holds no position in resources, currency and fixed income and hence is overweight financials and underweight fixed income relative to classic MVO, and is overweight financials and industrial relative to EW approach.

The HS-FP has a more erratic asset allocation profile in comparison to the classic MVO and naturally EW models which is not unexpected considering this approach doesn't follow a historical expected mean or heuristic approach to asset allocation, and asset class distributions are conditioned by time and state variables.  

\begin{figure}[h]
\centering
\includegraphics[width=0.5\textwidth]{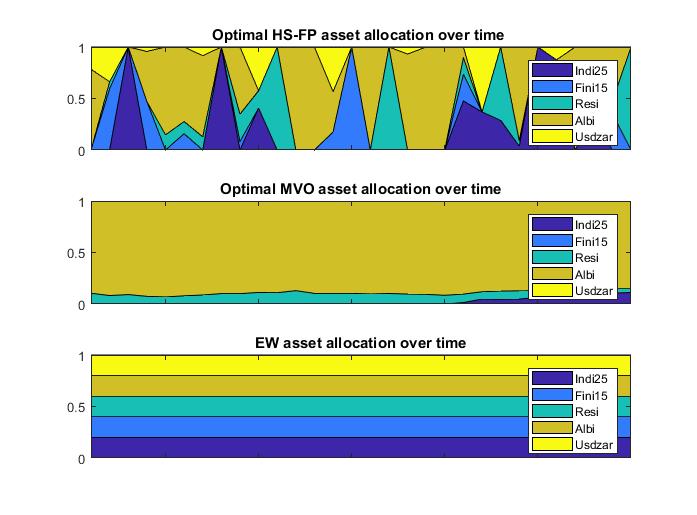}
\caption{Optimal Asset Allocation: The out of sample optimal allocation to each asset class over time is shown for the three portfolio construction methods considered; the Historical Simulation with Flexible Probabilities (HS-FP), Mean Variance Optimization (MVO) and Equal Weighting (EW) respectively.}
\label{fig:fig6}
\end{figure}

\subsubsection{Comparison of out-of-sample statistics} \label{sec:out-stat}

In this subsection we compare the out-of-sample performance, risk and turnover statistics of HS-FP to the benchmark models.

\FloatBarrier
\begin{table}[H]
\begin{center}
\begin{tabular}{|p{4cm}|p{1.5cm}|p{1.5cm}|p{1.5cm}|}
\hline
Approaches & Ann. Return & Ann. Volatility & Sharpe Ratio \\
 \hline 
HS-FP  & 11.16\% & 12.12\% & 0.32 \\
MVO  & 8.27\% & 6.26\% & 0.16 \\
EW  & 8.13\% & 10.91\% & 0.08 \\
\hline
\end{tabular}
\caption{Asset Allocation Models Summary Statistics I: Popular summary statistics for the three portfolio construction methodologies covered in this paper are considered, Historical Simulation with Flexible Probabilities (HS-FP), Mean Variance Optimal (MVO) and Equally Weighted (EW) respectively. The gross annualised return, annualised volatility, and annualised Sharpe ratio summary statistics are provided for the out of sample period of March 2003 until December 2017.}
\end{center}
\label{table:Table8}  
\end{table}

\begin{table}[H]
\begin{center}
\begin{tabular}{|p{4cm}|p{1.5cm}|p{1.5cm}|p{1.5cm}|}
\hline
Approaches & Max. Drawdown & CVaR & Turnover\\
\hline
HS-FP   & 21.20\% & 9.84\% & 10.93 \\
MVO  & 10.63\% & 4.01\% & 0.14 \\
EW  & 34.82\% & 7.98\% & 0.57 \\
\hline
\end{tabular}
\caption{Asset Allocation Models Summary Statistics II: Summary statistics for the three portfolio construction methodologies covered in this paper are considered, Historical Simulation with Flexible Probabilities (HS-FP), Mean Variance Optimal (MVO) and Equally Weighted (EW) respectively. Summary statistics such as maximum drawdown, conditional value at risk and average monthly turnover are calculated. These statistics are calculated as they look into historical rare events and assess the implication for turnover from using different portfolio construction methodologies. These summary statistics are shown for the March 2003 - Dec 2017 out of sample period.}
\end{center}
\label{table:Table9}  
\end{table}
\FloatBarrier

The best performing allocation model is the HS-FP approach in terms of annualised gross returns. However, the HS-FP model performs worse overall in terms of risk, it has higher volatility and CVaR than both benchmark models and higher maximum drawdown than classic MVO. It also has a higher turnover than both benchmark models which is unsurprising considering its erratic optimal asset allocation as shown in Figure \ref{fig:fig6}. Finally in terms of risk adjusted returns the Sharpe ratio is the highest for the HS-FP approach and is 2 and 4 times greater than the Sharpe ratio of classic MVO and EW respectively. From these results and based on the parameter assumptions it could be deduced that over the entire analysis period the HS-FP model has better risk adjusted returns than the benchmark models.
       
\subsubsection{Comparison of out of sample rolling returns}\label{sec:oos-rets}

In the previous subsection we see HS-FP model has better risk adjusted returns than the benchmark models. This was based on summary stats over the entire period, there  could be  periods of time historically where the HS-FP significantly under/over performed the benchmark models. In this section we look closer into the rolling returns of the HS-FP and Benchmark model to see whether there are notable periods with significant deviations from long term experience. We do this by observing cumulative returns and annualised returns relative to benchmark models.

\FloatBarrier
\begin{figure}[h]
\centering
\includegraphics[width=0.5\textwidth]{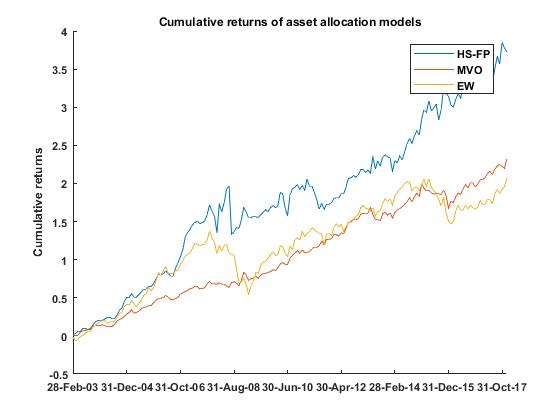}
\caption{Cumulative Return for Asset Allocation Models: A comparison of the out of sample cumulative returns for the Historical Simulation with Flexible Probabilities (HS-FP), Mean Variance Optimal (MVO) and Equally Weighted (EW) asset allocation models.}
\label{fig:fig9}
\end{figure}

\begin{figure}[h]
\centering
\includegraphics[width=0.5\textwidth]{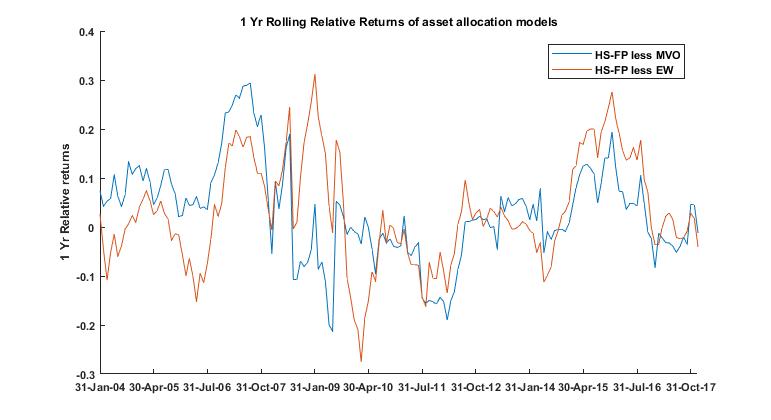}
\caption{Rolling Relative Returns for Asset Allocation Models: A comparison of the 1 Year out of sample relative returns for the Historical Simulation with Flexible Probabilities (HS-FP), Mean Variance Optimal (MVO) and Equally Weighted (EW) asset allocation models.}
\label{fig:fig10}
\end{figure}
\FloatBarrier

Based on the cumulative graph it appears that on a return basis the HS-FP model is far superior to benchmark models. When the relative return graph is considered we see periods where the HS-FP model significantly under and outperformed benchmark models. When performance of HS-FP model is compared to Classic MVO benchmark model we see that from 2003-2008 HS-FP outperformed the benchmark, then from 2008-2012 HS-FP struggled against benchmark and bounced back up recently from 2012-2017. In comparison to EW benchmark model, HS-FP model struggled initially from 2003-2006, outperformed from 2006-2009, then underperformed for the next 3 years and strongly outperformed from 2014 till recently.\\ 
During the credit crisis (2007-2008), the HS-FP model recorded an annualised return of $8.1\%$ outperforming both the classic MVO and EW benchmark models which recorded $7.7\%$ and $-7.5\%$ respectively. The major contributor of this out-performance was due to the HS-FP model having a greater overweight relative to the benchmark models on asset classes that performed better during the Global Financial Crisis (GFC). In terms of risk, the HS-FP model was more volatile during this period recording an annualised volatility of $23\%$ while classic MVO and EW recorded lower volatilities of $7\%$ and $14\%$ respectively. The HS-FP model had the highest Sharpe ratio during the GFC, recording an annualised risk adjusted return of -0.08 while classic MVO and EW models recorded lower Sharpe ratios of -0.32 and -1.28 respectively. The risk free rate during this period was $10\%$.\\
This relative performance suggests the cyclical nature of HS-FP model and flags the potential of backtest overfitting due to starting period assumption and hence under estimation of out-of-sample generalization errors. 

\subsection{Comparison with Balanced Funds}

It is also important to consider how the HS-FP and benchmark models fared in comparison to South African balanced funds. We consider the average performance of balanced funds from the ASISA South African (SA) Multi Asset(MA) Flexible category and the ASISA Global Multi Asset(MA) Flexible category over the past 5 years (Jan 2013-Dec 2017). The risk free rate over this period was $6.35\%$. 

\begin{table}[H]
\begin{center}
\begin{tabular}{|p{4cm}|p{1.5cm}|p{1.5cm}|p{1.5cm}|}
\hline
Approaches & Ann. Return & Ann. Volatility & Sharpe Ratio \\
 \hline 
HS-FP  & 8.23\% & 9.98\% & 0.19 \\
MVO & 5.07\% & 7.04\% & -0.18 \\
EW & 2.72\% & 9.09\% & -0.40 \\
SA MA Flexible  & 10.16\% & 6.72\% & 0.57 \\
Global MA Flexible & 14.64\% & 12.92\% & 0.54 \\
\hline
\end{tabular}
\caption{Summary Statistics for Asset Allocation Models relative to Balanced Funds: Popular summary statistics such as annualised return, annualised volatility and annualised Sharpe ratio for the Historical Simulation with Flexible Probabilities (HS-FP), Mean Variance Optimal (MVO) and Equally Weighted (EW) asset allocation models are compared to South African and Global Balanced funds over the period Jan 2013-Dec 2017.}
\end{center}
\label{table:Table99}  
\end{table}

In Table \ref{table:Table99} we observe that over the past 5 years, the HS-FP model recorded an annualised performance of $8.23\%$ and underperformed both the average of the SA MA Flexible category and Global MA Flexible category which recorded $10.16\%$ and $14.64\%$ respectively. In terms of risk, the volatility of HS-FP model is higher than that of the SA MA Flexible category but lower than that of the Global MA Flexible category. Finally in terms of the Sharpe ratio, the HS-FP has a lower Sharpe ratio than both SA MA Flexible category and Global MA Flexible category. A potential reason for this underperformance could be the lack of offshore asset classes in the asset class universe for the HS-FP and benchmark models. This suggests that there could have been a benefit to performance, risk and diversification for the HS-FP model and benchmark models from including offshore assets. Over the past 5 years, the HS-FP model maintains its outperformance relative to the classic MVO and EW benchmark models. 

\section{Robustness of out of sample results} \label{sec:stats}

The results discussed so far in Section \ref{sec:analysis} is one potential historical path based on initial assumption and parameter steps applied. In this section we analyse the robustness of the out-of-sample returns discussed in the previous section. We follow the backtesting protocol suggested in \cite{R2018} discussing assumptions and parameter steps that may lead to over-fitting and hence under estimating the out-of-sample generalization errors. We consider measures that assess performance inflation due to non-normality of asset class returns and due to selection bias under multiple testing. We also assess the legitimacy and consistency of backtest result when assumptions are varied. The Probabilistic Sharpe Ratio introduced in \cite{R2012}, Deflated Sharpe Ratio introduced in \cite{RR2014} and the Probability of Backtest Overfitting as suggested by Bailey et al \cite{R2016} are calculated and discussed in this section.

%The reason for this, as several studies have demonstrated, is the inflationary effect of short samples and samples drawn from non-Normal returns distributions. We refer the interested reader to Lo [2002], Mertens [2002], López de Prado and Peijan [2004], Ingersoll et al. [2007] for a discussion.

\subsection{Transaction cost}\label{sec:tc}

The results that were discussed in \ref{sec:analysis} where based on transaction cost assumed to be zero. In reality implementing an asset allocation strategy using ETF's or futures contracts have a cost associated with them. In this section we analyse the impact of transaction costs on the results previously found.  

\begin{table}[H]
\begin{center}
\begin{tabular}{|p{1.5cm}|p{1.1cm}|p{1.1cm}|p{1.1cm}|p{1.1cm}|p{1.1cm}|p{1.1cm}|}
\hline
 \multicolumn{7}{|c|}{Transaction cost = 0} \\
\hline
 Allocation Model & Ann. Return & Ann. Vol. & SR & Max DD & Cvar & Turn over\\
 \hline
 HS-FP  & 11.1\% & 12.1\% & 0.32 & 21.2\% & 9.8\% & 10.9\% \\
 MVO  & 8.2\% & 6.3\% & 0.16 & 10.6\% & 4.0\% & 0.1\% \\
 EW & 8.1\% & 10.9\% & 0.08 & 34.8\% & 7.9\% & 0.5\% \\
 \hline
\end{tabular}
\caption{Summary Statistics for Transaction Cost = 0 bps: Summary statistics such as annualised return, annualised volatility, annualised Sharpe ratio, maximum drawdown, conditional value at risk and average monthly turnover are shown for the three asset allocation models; when the transaction cost is set to 0 bps, and the out of sample performance is observed over the March 2003 - Dec 2017 period.}
\end{center}
\label{table:Table10}  
\end{table}

\begin{table}[H]
\begin{center}
\begin{tabular}{|p{1.5cm}|p{1.1cm}|p{1.1cm}|p{1.1cm}|p{1.1cm}|p{1.1cm}|p{1.1cm}|}
\hline
 \multicolumn{7}{|c|}{Transaction cost = 20 bps} \\
\hline
 Allocation Model & Ann. Return & Ann. Vol. & SR & Max DD & Cvar & Turn over\\
 \hline
 HS-FP  & 10.9\% & 12.1\% & 0.30 & 21.4\% & 9.9\% & 10.9\% \\
 MVO  & 8.2\% & 6.3\% & 0.16 & 10.6\% & 4.0\% & 0.1\% \\
 EW & 8.1\% & 10.9\% & 0.08 & 34.8\% & 7.9\% & 0.5\% \\
 \hline
\end{tabular}
\caption{Summary Statistics for Transaction Cost = 20 bps: Summary statistics such as annualised return, annualised volatility, annualised Sharpe ratio, maximum drawdown, conditional value at risk and average monthly turnover are shown for the three asset allocation models; when the transaction cost is set to 20 bps, and the out of sample performance is observed over the March 2003 - Dec 2017 period.}
\end{center}
\label{table:Table11}  
\end{table}

\begin{table}[H]
\begin{center}
\begin{tabular}{|p{1.5cm}|p{1.1cm}|p{1.1cm}|p{1.1cm}|p{1.1cm}|p{1.1cm}|p{1.1cm}|}
\hline
 \multicolumn{7}{|c|}{Transaction cost = 30 bps} \\
\hline
 Allocation Model & Ann. Return & Ann. Vol. & SR & Max DD & Cvar & Turn over\\
 \hline
 HS-FP  & 10.7\% & 12.1\% & 0.28 & 21.5\% & 10.0\% & 10.9\% \\
 MVO  & 8.2\% & 6.3\% & 0.16 & 10.6\% & 4.0\% & 0.1\% \\
 EW & 8.1\% & 10.9\% & 0.07 & 34.8\% & 7.9\% & 0.5\% \\
 \hline
\end{tabular}
\caption{Summary Statistics for Transaction Cost = 30 bps: Summary statistics such as annualised return, annualised volatility, annualised Sharpe ratio, maximum drawdown, conditional value at risk and average monthly turnover are shown for the three asset allocation models; when the transaction cost is set to 30 bps, and the out of sample performance is observed over the March 2003 - Dec 2017 period.}
\end{center}
\label{table:Table12}  
\end{table}

\begin{table}[H]
\begin{center}
\begin{tabular}{|p{1.5cm}|p{1.1cm}|p{1.1cm}|p{1.1cm}|p{1.1cm}|p{1.1cm}|p{1.1cm}|}
\hline
 \multicolumn{7}{|c|}{Transaction Cost = 50 bps} \\
\hline
 Allocation Model & Ann. Return & Ann. Vol. & SR & Max DD & Cvar & Turn over\\
 \hline
 HS-FP  & 10.5\% & 12.2\% & 0.26 & 21.7\% & 10.0\% & 10.9\% \\
 MVO  & 8.2\% & 6.3\% & 0.16 & 10.6\% & 4.0\% & 0.1\% \\
 EW & 8.1\% & 10.9\% & 0.07 & 34.8\% & 7.9\% & 0.5\% \\
 \hline
\end{tabular}
\caption{Summary Statistics for Transaction Cost = 50 bps: Summary statistics such as annualised return, annualised volatility, annualised Sharpe ratio, maximum drawdown, conditional value at risk and average monthly turnover are shown for the three asset allocation models; when the transaction cost is set to 50 bps, and the out of sample performance is observed over the March 2003 - Dec 2017 period.}
\end{center}
\label{table:Table13}  
\end{table}

Overall, we see the conclusions drawn in Section \ref{sec:analysis} do not change, the HS-FP approach still has the highest Sharpe ratio relative to benchmark strategies even at transaction cost set at 50 bps. The Sharpe ratio does fall from 0.32 in the case of 0 bps transaction cost to 0.26 in the case of 50 bps transaction cost.

\subsection{Training window} \label{sec:twindow}

In Section \ref{sec:analysis} we used 5 years as a initial training period we made this assumption so that we could assess the out of sample performance of the HS-FP model in the Global Financial Crisis (GFC). In this subsection we discuss the impact of altering the initial period or training window on the results observed so far.

\begin{table}[H]
\begin{center}
\begin{tabular}{|p{1.5cm}|p{1.1cm}|p{1.1cm}|p{1.1cm}|p{1.1cm}|p{1.1cm}|p{1.1cm}|}
\hline
 \multicolumn{7}{|c|}{Training window = 5 years} \\
\hline
 Allocation Model & Ann. Return & Ann. Vol. & SR & Max DD & Cvar & Turn over\\
 \hline
HS-FP  & 11.1\% & 12.1\% & 0.32 & 21.2\% & 9.8\% & 10.9\% \\
MVO  & 8.2\% & 6.3\% & 0.16 & 10.6\% & 4.0\% & 0.1\% \\
EW & 8.1\% & 10.9\% & 0.08 & 34.8\% & 7.9\% & 0.5\% \\
 \hline
\end{tabular}
\caption{Summary Statistics for Training Window = 5 years: Summary statistics such as annualised return, annualised volatility, annualised Sharpe ratio, maximum drawdown, conditional value at risk and average monthly turnover are shown for the three asset allocation models; when the initial training window is set to 5 years, and the out of sample returns are observed over the March 2003 - Dec 2017 period.} 
\end{center}
\label{table:Table14}
\end{table}

\begin{table}[H]
\begin{center}
\begin{tabular}{|p{1.5cm}|p{1.1cm}|p{1.1cm}|p{1.1cm}|p{1.1cm}|p{1.1cm}|p{1.1cm}|}
\hline
 \multicolumn{7}{|c|}{Training window = 8 years} \\
\hline
 Allocation Model & Ann. Return & Ann. Vol. & SR & Max DD & Cvar & Turn over\\
 \hline
HS-FP  & 8.9\% & 13.0\% & 0.14 & 21.2\% & 10.9\% & 11.1\% \\
MVO  & 6.9\% & 6.5\% & -0.08 & 10.6\% & 4.1\% & 0.1\% \\
EW & 4.9\% & 11.0\% & -0.18 & 34.8\% & 8.3\% & 0.5\% \\
 \hline
\end{tabular}
\caption{Summary Statistics for Training Window = 8 years: Summary statistics such as annualised return, annualised volatility, annualised Sharpe ratio, maximum drawdown, conditional value at risk and average monthly turnover are shown for the three asset allocation models; when the initial training window is set to 8 years, and the out of sample returns are observed over the March 2006 - Dec 2017 period.} 
\end{center}
\label{table:Table15} 
\end{table}

\begin{table}[H]
\begin{center}
\begin{tabular}{|p{1.5cm}|p{1.1cm}|p{1.1cm}|p{1.1cm}|p{1.1cm}|p{1.1cm}|p{1.1cm}|}
\hline
 \multicolumn{7}{|c|}{Training window = 10 years} \\
\hline
 Allocation Model & Ann. Return & Ann. Vol. & SR & Max DD & Cvar & Turn over\\
 \hline
 HS-FP  & 7.9\% & 13.2\% & 0.09 & 21.2\% & 11.1\% & 10.3\% \\
 MVO  & 7.1\% & 6.8\% & 0.07 & 10.6\% & 4.4\% & 0.1\% \\
 EW & 4.5\% & 11.2\% & -0.19 & 29.6\% & 8.6\% & 0.5\% \\
 \hline
\end{tabular}
\caption{Summary Statistics for Training Window = 10 years: Summary statistics such as annualised return, annualised volatility, annualised Sharpe ratio, maximum drawdown, conditional value at risk and average monthly turnover are shown for the three asset allocation models; when the initial training window is set to 10 years, and the out of sample returns are observed over the March 2008 - Dec 2017 period.} 
\end{center}
\label{table:Table16} 
\end{table}

\begin{table}[H]
\begin{center}
\begin{tabular}{|p{1.5cm}|p{1.1cm}|p{1.1cm}|p{1.1cm}|p{1.1cm}|p{1.1cm}|p{1.1cm}|}
\hline
 \multicolumn{7}{|c|}{Training window = 12 years} \\
\hline
 Allocation Model & Ann. Return & Ann. Vol. & SR & Max DD & Cvar & Turn over\\
 \hline
 HS-FP  & 7.6\% & 10.4\% & 0.14 & 13.0\% & 5.9\% & 10.3\% \\
 Classic MVO  & 7.5\% & 6.6\% & 0.22 & 10.6\% & 4.4\% & 0.1\% \\
 EW & 5.5\% & 9.7\% & -0.06 & 19.5\% & 6.1\% & 0.5\% \\
 \hline
\end{tabular}
\caption{Summary Statistics for Training Window = 12 years: Summary statistics such as annualised return, annualised volatility, annualised Sharpe ratio, maximum drawdown, conditional value at risk and average monthly turnover are shown for the three asset allocation models; when the initial training window is set to 12 years, and the out of sample returns are observed over the March 2010 - Dec 2017 period.}
\end{center}
\label{table:Table17} 
\end{table}

We do see changes in the results drawn in \ref{sec:analysis} when the training window is varied, the HS-FP model maintains a higher Sharpe ratio relative to EW benchmark at different training windows but underperforms the classic MVO at training window 12 years. 

\subsection{Probabilistic Sharpe Ratio (PSR)}\label{sec:PSR}

The Probabilistic Sharpe Ratio (PSR) \cite{R2016,AA2019} is a uncertainty-adjusted investment skill metric that measures the probability that a strategy exceeds a benchmark threshold in the presence of non-normal returns. We consider the PSR for the following reasons: to assess whether the HS-FP consistently outperforms the MVO and EW benchmark models once corrected for performance inflation due to non-normality of asset class returns. The PSR is the probability that the true SR is above a given rejection threshold. The rejection threshold is set as the Sharpe ratio of the Benchmark models specified in Section \ref{sec:bench}. The PSR takes into account the sample length, skewness and kurtosis of the returns’ distribution. 

%For further details regarding the calculation of PSR see Appendix \ref{sec:PSRcalc}. 

In Table \ref{table:Table8} we see that the HS-FP point estimate Sharpe ratio is 2 times and 4 times greater than the classic MVO and EW respectively. Based on this result it could be tempting to believe that the HS-FP Sharpe ratio is significantly greater than the benchmark models. However, when the PSR is calculated it is seen that the HS-FP model outperforms EW benchmark at 20 percent significance level and outperforms the MVO benchmark at 30 percent significance level. At 5 percent significance level the HS-FP performance is not statistically significant relative to both benchmark model. 

\begin{table}[H]
\begin{center}
\begin{tabular}{|p{4cm}|p{1.5cm}|p{1.5cm}|p{1.5cm}|}
\hline
PSR Matrix & HS-FP & MVO & EW\\
\hline
 HS-FP  & 0.50 & 0.72 & 0.81\\
 MVO  & 0.27 & 0.50 & 0.62\\
 EW & 0.17 & 0.38 & 0.50\\
 \hline
\end{tabular}
\caption{PSR Matrix: The table represents the PSR matrix, the rows of the table represent the candidate strategy and the columns represent the benchmark strategy. The table should be interpreted as the probability that the candidate strategy SR exceeds the benchmark strategy SR.} 
\end{center}
\label{table:Table19} 
\end{table}

Additionally, when the Minimum Track Record Length (MinTRL) is calculated to see appropriate track record length, needed, to reject null hypothesis that: HS-FP's Sharpe ratio displays no skill beyond MVO and EW Sharpe ratio thresholds, at a 95 percent confidence level. We see the MinTRL is 1404 and 624 monthly observations or 117 and 52 years of track record is required. So the 179 months of out-of-sample performance we have currently is too short and does not meet the MinTRL required to reject the null hypothesis that HS-FP's Sharpe ratio is below the benchmark models threshold SR's at a 95$\%$ confidence level.\\ 

In this subsection we assess the significance of the single out-of-sample historic simulation or single trial we calculated in Section \ref{sec:analysis}. The PSR shows that the Sharpe ratio of the HS-FP model is not statistically significant relative to benchmark models additionally the MinTRL shows that the number of out of sample returns is too short to claim significance. These results indicate reasons to doubt the significance of the HS-FP’s Sharpe ratio relative to benchmark models as seen in Section \ref{sec:analysis}.

\subsection{Probability of Backtest Overfitting (PBO)}\label{sec:PBO}

The probability of backtest overfitting is the non-null probability that a strategy with optimal performance In Sample (IS) ranks below the median Out Of Sample (OOS) \cite{R2016,AA2019}. We consider the probability of backtest overfitting in this paper to consider whether the results shown constitutes a legitimate empirical finding which holds true under varied assumptions and parameters.

We estimate the probability of backtest overfitting through Combinatorially Symmetric Cross Validation(CSCV). Similar to \cite{R2016} the following overfit statistics are calculated: 

\begin{enumerate}
\item Probability of Backtest Overfitting (PBO) is the probability that the model configuration selected as optimal IS will underperform the median of the N model configurations OOS.
\item Performance degradation which determines to what extent greater performance IS leads to lower performance OOS.
\item Probability of loss which is the probability that the model selected as optimal IS will deliver a loss OOS.
\item Stochastic dominance determines whether the procedure used to select a strategy IS is preferable to randomly choosing one model configuration among the N alternatives.
\end{enumerate}
We analyse the PBO through CSCV approach by considering 5 parameters: leeway, rebalancing frequency and state variable data transformation parameters i.e. prior half life, fast half-life and slow half-life. In the main results sections we had assumed leeway = 0.1, rebalancing frequency = semi-annually, prior half life = 5 years, fast half-life = 3 months, slow half-life = 12 months. In the probability of backtest overfitting analysis we explore the following ranges for these parameters: leeway= 0.1,0.2,0.3, rebalancing frequency= 1,2,3,...,6,...,12 (monthly, bi-monthly, quarterly,.., half-yearly,...,annually), prior half life = 5,6,7,8 years, fast half life = 3,6,9,12 months and slow half life=12,18,24,36 months. The parameter combination of the before mentioned results in a 5 dimensional mesh of 2304 elements.

We estimate PBO using the CSCV procedure and set the threshold Sharpe Ratio (SR) as 0. 
 
\FloatBarrier
\begin{figure}[h]
\centering
\includegraphics[width=0.5\textwidth]{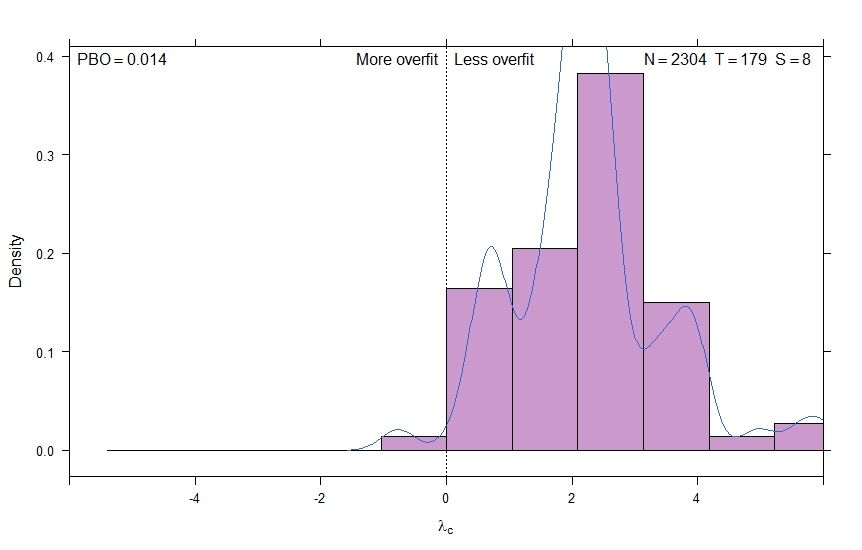}
\caption{Shows the distribution of Logits and Probability of Backtest Overfitting (PBO) for the HS-FP model, the Logits is the logarithm of odds and in our analysis the odds represents the odds that the optimal strategy chosen IS happens to underperform OOS. The logits are then used to calculate the PBO. When the distribution of Logits is centered in a significantly positive value with left tail marginally covering the region of negative logits this implies the backtest results are conducive to good OOS results.}
\label{fig:fig12}
\end{figure}

\begin{figure}[h]
\centering
\includegraphics[width=0.5\textwidth]{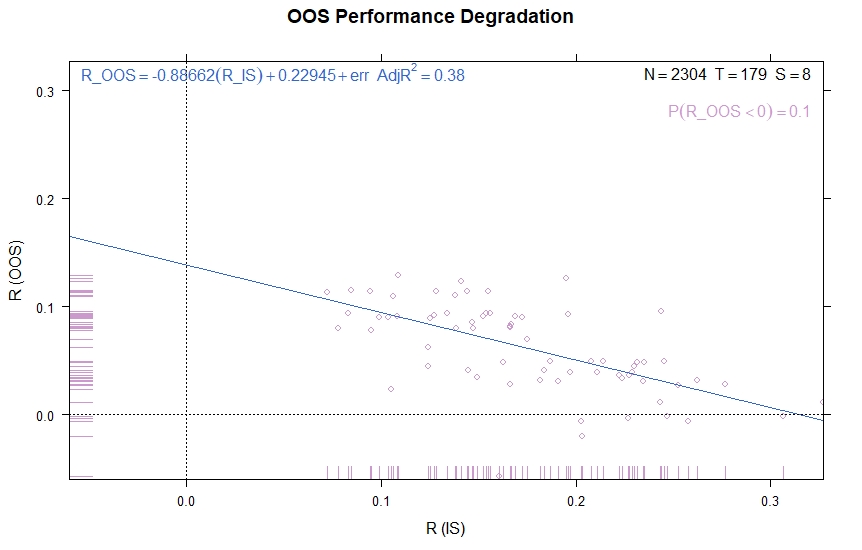}
\caption{Shows the pairs of (SR IS,SR OOS) for optimal model configuration or the performance degradation associated with HS-FP model backtest and the OOS probability of loss.}
\label{fig:fig13}
\end{figure}
\FloatBarrier

\begin{figure}[h]
\centering
\includegraphics[width=0.5\textwidth]{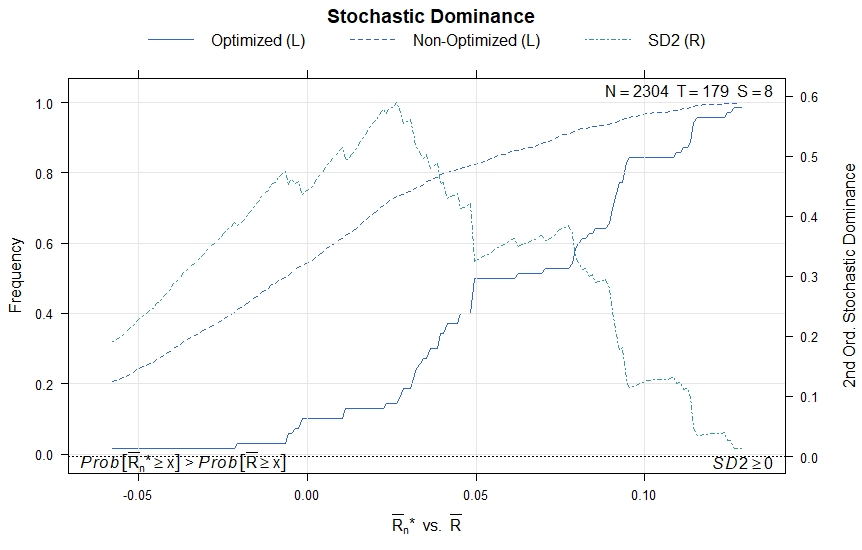}
\caption{Shows the stochastic dominance of HS-FP with the cumulative distribution function of best OOS SR across all combinations (optimized) and OOS SR (non-optimized), as well as the second order stochastic dominance (SD2) for every OOS SR.}
\label{fig:fig14}
\end{figure}

Figure ~\ref{fig:fig12} plots the distribution of logits, the logits is the logarithm of odds and in our analysis the odds represents the odds that the optimal strategy chosen IS happens to underperform OOS. The distribution of logits centered in a significantly positive value with left tail marginally covering the region of negative logits implies backtest results are conducive to good OOS results. In our analysis the distribution of logits is centered around a positive value with left tail marginally covering negative logits thus implying the backtest results are conducive to good OOS results. Using the logits we calculate PBO of approximately 1.5$\%$.

The regression line that goes through the pairs of SR IS and SR OOS as seen in Figure ~\ref{fig:fig13} has a negative slope of -0.88, this implies that the higher the SR is IS the lower the SR is OOS. Indicating seeking the optimal performance at some point becomes detrimental. Figure ~\ref{fig:fig13} also shows that approximately 10$\%$ of HS-FP SR OOS are negative despite all SR IS being positive and ranging approximately between 0.07 and 0.32.

Finally, we look at the stochastic dominance, to determine whether the distribution of the best performing OOS SR across all combinations stochastically dominates the distribution of all OOS SR. If this dominance is not seen then it would present strong evidence that selecting by best SR does not provide consistently better OOS results than a random strategy selection criterion. In our analysis Figure ~\ref{fig:fig14} indicates that selecting by best SR added value, since the distribution of OOS performance for the best performing OOS SR dominates the overall distribution of OOS SR. This can be seen by the fact that for every level of OOS SR the proportion of optimized model configurations is lesser than the proportion of non-optimized, thus the probabilistic mass of the optimized model is shifted to the right of the non-optimized model. Second-order dominance as seen in Figure ~\ref{fig:fig14} on y2-axis implies that the distribution does not dominate but first-order stochastic dominance is a sufficient condition for second-order stochastic dominance.

\subsection{Deflated Sharpe Ratio (DSR)}\label{sec:DSR}

The Deflated Sharpe Ratio (DSR) is a uncertainty-adjusted investment skill metric that measures the probability that a strategy exceeds a benchmark threshold in the presence of non-normal returns and selection bias under multiple testing. Performance inflation can occur due to non-normality of asset class returns and due to selection bias under multiple testing. In Section \ref{sec:PSR} we assess the probability of performance inflation due to only non-normality of asset class returns. In this section we assess the probability of performance inflation due to non-normality of asset class returns and selection bias under multiple testing. We calculate the DSR to assess the legitimacy of the backtest results shown in Section \ref{sec:analysis}. The DSR is a PSR where the rejection threshold is adjusted to reflect the multiplicity of trials. We calculate the DSR using the method suggested in the paper \cite{AA2019}. This method involves using optimal K-means cluster analysis to reduce the number of total trials (N) to approximate the number of independent trials (K), deriving optimal weights for each of the strategies within the independent trials by using inverse variance optimization, then calculating the aggregate Sharpe ratio for each of the independent trials and finally calculating the variance of these K Sharpe ratios. The number of total trials in our case is N=2304 see Section \ref{sec:PBO}, from the recursive cluster analysis the number of independent trials is calculated to be K=9, the length of observations is T=179 months, the observed $SR^{*}$ equals 0.09 and the user defined benchmark SR adjusted for multiplicity of trials $\hat{SR}$ equals 0.11, this results in a DSR of 0.40. This DSR implies there is 40$\%$ chance that the true SR associated with this strategy is greater than zero. 

%For further details regarding the calculation of DSR see Appendix \ref{sec:DSRcalc}.

%sort latex and formulae in appendix

\section{Conclusions} \label{sec:conclude}

Here we have implemented a systematic asset allocation model which uses the Historical Simulation with Flexible Probabilities approach. This is a non-parametric approach to asset allocation which considers future asset behavior to be conditional on different state variables and derives a forward looking distribution that is consistent with investor views but remains as close as possible to the prior distribution. 

Key benefits of this framework are that: First, it allows for time-varying expected returns and covariances, Second, this framework allows one to combine multiple state variables which can help reduce the noise of individual state variables and potentially mitigate the risk of over-fitting and model miss-specification. Here we have considered state variables ranging from macroeconomic, trend and risk based indicators (see Table \ref{table:Table3}, Section \ref{sec:single-var} and Section \ref{sec:multi-var}).

Although the key contribution of this paper is to consider the potential effectiveness of a simple HS-FP implementation in the context of South African markets, the work can be of interest to a more general audience as it links the HS-FP framework to an explicit benchmark based test, and a quantitative reflection on the perils of backtest overfitting (see Section \ref{sec:stats}).

The main out-of-sample results are over the period 2003-2017, where the HS-FP model combines all state variables and results in a Sharpe ratio that exceeds both classic MVO and EW benchmark models (see Section \ref{sec:out-stat}, Table \ref{table:Table8}). These results hold true at varied transaction costs, even as high as 50 bps (see Section \ref{sec:tc}, Table \ref{table:Table13}). 

However, these results are inconsistent when training windows are varied (see Section \ref{sec:twindow}), the point estimate Sharpe ratio is inflated (see Table \ref{table:Table19}, Section \ref{sec:PSR} and Section \ref{sec:DSR}), and the track record length fails to be above the minimum track record length required for statistical significance (see Section \ref{sec:PSR}). 

We are able to show low probability that the backtested returns from HS-FP model under varied parameters could have been overfitted (see Section \ref{sec:PBO}); but this does not provide compelling evidence that this approach can be naively used for asset management decision making, neither smart beta type indexation nor even fully automated quasi-active management, at least in its current implementation.

Can this type of idea deliver in a real way? We think it can, and we think that we have provided some evidence to support the claim in the context of the South African market use case. Can this protect against bad events? It is almost certain that it cannot protect against all bad events. However, time-scales are everything in finance, what we have shown is that the strategy did relatively well through the Global Financial Crises on an out-of-sample basis (see Section \ref{sec:oos-rets}), even though there is an increase in portfolio turnover. There does appear to be a risk benefit to the higher levels of rebalancing required to implement the strategy after costs (see Table \ref{table:Table13}) and over the recent past (see Table \ref{table:Table99}). However this is only on a Sharpe Ratio basis, and not in terms of the maximum drawdown nor CVaR (See Table \ref{table:Table9}). It is conceivable that a relatively shorter and more adaptive time-horizon where more cross-sectional data is quantitatively included in investment decision making process may have an effective risk-return advantage.  

In future research we hope to incorporate transaction costs directly in the portfolio construction process itself, investigate the performance of the HS-FP model across broader asset classes and regions, and consider a broader set of state variables that could increase the breadth of information used in order to mitigate window dependencies, improve the forward looking adaptability and predictive power, while controlling the rebalancing variance of the model implementation.  

\section{Acknowledgments}

The authors would like to thank Andrew Paskaramoothy, Igor Rodionov, Dieter Hendricks, Raphael Nkomo and Lara Dalmeyer for discussions relating to the work. Particular thanks are extended to Graham Barr for thoughtful comments and critique. The authors have declared that no competing interests exist and that any views or opinions presented are solely those of the authors and do not necessarily represent those of STANLIB nor the University of Cape Town. TG acknowledges research support from the University of Cape Town (fund 459282).

\section{The Algorithms} \label{sec:alg}
\FloatBarrier
\subsection{Exponential Decay} \label{ssec:ewma}
\begin{algorithm}
\begin{algorithmic}
\Function{Exponential decay}{$\bar{t}$,$\tau_{_{HL}}$}
 {\\ \begin{enumerate} 
\item Input:\begin{enumerate}
\item Length of exponential decay probability, $\bar{t}$
\item Half life parameter, $\tau_{_{HL}}$
\end{enumerate}
\item Output: \begin{enumerate} 
\item exponential decay probability, $p^{\exp}$
\end{enumerate}
\item compute probability, $\{p_{t}^{\exp}\}_{t = 1}^{\bar{t}} = \bigg\{e^{-\frac{\ln{2}}{\tau_{_{HL}}}|\bar{t}-t|}\bigg\}_{t = 1}^{\bar{t}}$
\item rescale, $\{p_t|\tau_{_{HL}}\}_{t=1}^{\bar{t}} = \{(p_t|\tau_{_{HL}})/\sum_{s=1}^{t}(p_s|\tau_{_{HL}})\}_{t=1}^{\bar{t}}$
\end{enumerate}}
\EndFunction
\end{algorithmic} 
\caption{Exponential decay} \label{alg:EWMA}
\end{algorithm}

\subsection{Crisp Probability} \label{ssec:crisp}
\begin{algorithm}[H]
\begin{algorithmic}
\Function{Crisp Probability}{$z_t$, $z_{t}^{*}$,$\alpha$}
 {\\ \begin{enumerate} 
\item Input:\begin{enumerate}
\item state variable, $\{z_t\}_{t=1}^{\bar{t}}$
\item target values, $\{z_{t}^{*}\}_{t=1}^{\bar{t}}$
\item Leeway, $\alpha$
\end{enumerate}
\item Output: \begin{enumerate} 
\item crisp probabilities, $\{p|\mathcal{R}(z_{k}^{*})\}_{k=1}^{\bar{k}}$
\item lower bound, $\{\underaccent{\bar}{z}_k\}_{k=1}^{\bar{k}}$
\item upper bound, $\{\bar{z}_{k}\}_{k=1}^{\bar{k}}$
\end{enumerate}
\item sort state variable sample, $\{z_{t}^{\mathrm{sort}}\} = \mathrm{sort}(\{z_t\}_{t=1}^{\bar{t}})$
\item calculate sorted state variables cdf, $$\{F_{z}(z_{t}^{\mathrm{sort}})\}_{t=1}^{\bar{t}} = \mathrm{cdf} \bigg(\{z_t\}_{t=1}^{\bar{t}},\bigg\{\frac{1}{t}\bigg\}_{t=1}^{\bar{t}}\bigg)$$
\item evaluated at the target values, $\{F_{z}(z_{k}^{*})\}_{k=1}^{\bar{k}}$
\item lower quantile of state variable, $$z_{\min} = \mathrm{quantile}\bigg(\frac{\alpha}{2},\{z_t\}_{t=1}^{\bar{t}}\bigg)$$
\item upper quantile of state variable, $$z_{\max} = \mathrm{quantile}\bigg(1-\frac{\alpha}{2},\{z_t\}_{t=1}^{\bar{t}}\bigg)$$
\end{enumerate}}
\For {$k = 1,\ldots,k$} 
\If{$P\{Z_t > z_{k}^{*}<\alpha/2\}$ or $P\{Z_t \leq z_{k}^{*}<\alpha/2\}$}
\State{flatten cdf}
\EndIf
\If{$F_z(z_{k}^{*}) \geq 1-\alpha/2$}
\State{$F_z(z_{k}^{*} = 1-\alpha/2$}
\EndIf
\If{$F_z(z_{k}^{*}) \leq \alpha/2$}
\State{$F_z(z_{k}^{*}) = \alpha/2$}
\EndIf
\If{$z_{k}^{*} \leq z_{\min}$}
\State{$\underaccent{\bar}{z}_k = \min(\{z_t\}_{t=1}^{\bar{t}})$}
\State{$\bar{z}_k = \texttt{smooth\_quantile}(F_{z}(z_{k^{*}})+\alpha/2,\{z_t\}_{t=1}^{\bar{t}})$}
\EndIf
\If{$z_{k}^{*} \geq z_{\max}$}
\State{$\bar{z}_k = \texttt{smooth\_quantile}(F_{z}(z_{k^{*}})-\alpha/2,\{z_t\}_{t=1}^{\bar{t}})$}
\State{$\bar{z}_{k} = \max(\{z_t\}_{t=1}^{\bar{t}})$}
\Else
\State{$\bar{z}_k = \texttt{smooth\_quantile}(F_{z}(z_{k^{*}})-\alpha/2,\{z_t\}_{t=1}^{\bar{t}})$}
\State{$\bar{z}_k = \texttt{smooth\_quantile}(F_{z}(z_{k^{*}})+\alpha/2,\{z_t\}_{t=1}^{\bar{t}})$}
\EndIf
\begin{enumerate} 
\item[] $\{p_t|\mathcal{R}(z_{k}^{*})\}_{t=1}^{\bar{t}} = \bigg\{\textbf{1}_{\underaccent{\bar}{z}_k\leq z_t \leq \bar{z}_k}\bigg\}_{t=1}^{\bar{t}}$
\item[] $\{p_t|\mathcal{R}(z_{k}^{*})\}_{t=1}^{\bar{t}} = \{\{(p_t|\mathcal{R}(z_{k}^{*})\}_{k=1}^{\bar{k}})/\sum_{s=1}^{\bar{t}}(p_s|\mathcal{R}(z_{k}^{*})\}_{t=1}^{\bar{t}}$
\end{enumerate} 
\EndFor
\EndFunction
\end{algorithmic} [H]
\caption{Crisp Probability} \label{alg:crisp}
\end{algorithm}

\subsection{Time and State Conditioned Probability} \label{ssec:tscp}
\begin{algorithm}[H]
\begin{algorithmic}
\Function{time and state probability}{$z_t$, $z_{t}^{*}$,$\alpha$}
 {\\ \begin{enumerate} 
\item Input:\begin{enumerate}
\item state variable, $\{z_t\}_{t=1}^{\bar{t}}$
\item target values, $\{z_{t}^{*}\}_{t=1}^{\bar{t}}$
\item Leeway, $\alpha$
\end{enumerate}
\item Output: \begin{enumerate} 
\item time and state probability, $\tilde{p}$
\end{enumerate}
\item compute crisp probabilities, $\{p^{crisp(k)}\}_{k=1}^{\bar{k}} = \texttt{crisp\_fp}(\{z_{t}\}_{t=1}^{\bar{t}},\{z_{k}^{*}\}_{k=1}^{\bar{k}},\alpha)$
\end{enumerate}}
\For {$k = 1,\ldots,\bar{k}$} 
\begin{enumerate} 
\item[] $m|z_{k}^{*} = \sum_{t=1}^{\bar{t}}z_{t}p_{t}^{\mathrm{crisp}(k)}$ 
\item[] $s^2|z_{k}^{*} = \sum_{t=1}^{\bar{t}}z_{t}^{2}p_{t}^{\mathrm{crisp}(k)}-(m|z_{k}^{*})^{2}$ 
\item[] $\{a_{t}^{\mathrm{ineq}}\}_{t=1}^{\bar{t}} = \{z_{t}^{2}\}_{t=1}^{\bar{t}}$
\item[] $b^{\mathrm{ineq}} = (m|z_{k}^{*})^{2}+s^{2}|z_{k}^{*}$
\item[] $\{a_{t}^{\mathrm{eq}}\}_{t=1}^{\bar{t}} = \{z_{t}\}_{t=1}^{\bar{t}}$
\item[] $b^{\mathrm{eq}} = m|z_{k}^{*}$
\item[] $p|(z_{k}^{*}) = \texttt{min\_rel\_entropy\_sp}(p|\tau_{_{HL}},a^{\mathrm{ineq}},b^{\mathrm{ineq}},a^{\mathrm{eq}},b^{\mathrm{eq}})$
\end{enumerate} 
\EndFor
\EndFunction
\end{algorithmic} 
\caption{Time and State probability} \label{alg:tscp}
\end{algorithm}

\subsection{Effective Number of Scenarios} \label{ssec:ens}
\begin{algorithm}
\begin{algorithmic}
\Function{effective number of scenarios}{$p^{\exp}$, $\texttt{type\_ent}$,$\gamma$}
 {\\ \begin{enumerate} 
\item Input:\begin{enumerate}
\item exponential entropy, $p^{\exp}$
\item type of entropy, $\texttt{type\_ent}$
\item parameter of exponential entropy, $\gamma$
\end{enumerate}
\item Output: \begin{enumerate} 
\item effective number of scenarios, $\mathrm{ens}(p)$
\end{enumerate}
\If{$\texttt{type\_ent} = \exp~~~~or~~~\mathrm{none}$}
\State{$\mathrm{ens}(p) = e^{-\sum_{t=1}^{\bar{t}}}p_{t}\ln{p_t}$}
\Else
\State{$\mathrm{ens}_{\gamma}(p) = [\sum_{t=1}^{\bar{t}}(p_t)^{\gamma}]^{-1/(\gamma-1)}$}
\EndIf
\EndFunction
\end{enumerate}}
\end{algorithmic} 
\caption{Effective Number of Scenarios} \label{alg:ens}
\end{algorithm}

\subsection{Minimum Relative Entropy} \label{ssec:mre}
\begin{algorithm}[H]
\begin{algorithmic}
\Function{Minimum Relative Entropy}{$\mathrm{prior}$,$a^{\mathrm{ineq}}$,$b^{\mathrm{ineq}}$,$a^{\mathrm{eq}}$,$b^{\mathrm{eq}}$}
 {\\ \begin{enumerate} 
\item Input:\begin{enumerate}
\item prior probability, $\underaccent{\bar}{p}$
\item inequality constraints, $a^{\mathrm{ineq}}$,$b^{\mathrm{ineq}}$
\item equality constraints, $a^{\mathrm{eq}}$,$b^{\mathrm{eq}}$
\end{enumerate}
\item Output: \begin{enumerate} 
\item posterior probability, $\bar{p}$
\end{enumerate}
\item concatenate the inequality and equality constraints, $v = \begin{pmatrix}a^{ineq}\\a^{eq}\end{pmatrix},$ $\mu = \begin{pmatrix}b^{\mathrm{ineq}}\\b^{\mathrm{eq}}\end{pmatrix}$
\item normalize the constraints to avoid numerical problem,
\begin{enumerate}
\item $\hat{m}_v = \mean(\{v_{.j}\}_{j=1}^{\bar{j}})$
\item $\hat{s}_{v} = \std(\{v_{.j}\}_{j=1}^{\bar{j}})$
\item $\{v_{.j}\}_{j=1}^{\bar{j}} = \{(v_{.j}-\hat{m}_{v})./\hat{s}_{v}\}_{j=1}^{\bar{j}}$
\item $\mu = (\mu-\hat{m}_{v})./\hat{s}_v$
\end{enumerate}
\item for probability,
\begin{enumerate} 
\item $p(l) = \bigg\{\underaccent{\bar}{p}^{(j)} \times e^{l \times v_{.j}}/\sum_{k=1}^{\bar{j}\underaccent{\bar}{p}^{(k)} \times e^{l \times v_{.k}}}\bigg\}$
\item $h(l) = \ln{\sum_{j=1}^{\bar{j}}\underaccent{\bar}{p}^{(j)} \times e^{l \times (v_{.j}-\mu)}}$
\item $\nabla_{l}h(l) = \sum_{j=1}^{\bar{j}}[p(l)]_{j} \times (v_{.j}-\mu)$
\item $\nabla_{l}^{2}h(l) = \sum_{j=1}^{\bar{j}}[p(l)]_{j} \times (v_{.j}-\mu-\nabla_{l}h(l)) \times (v_{.j}-\mu-\nabla_{l}h(l))$
\end{enumerate}
\item compute,
\begin{enumerate}
\item $\theta = \mathrm{minimize}\Bigg(h(l),a^{ineq}=\begin{pmatrix}\mathcal{J} & 0\\0 & 0\end{pmatrix},b^{\mathrm{ineq}}=0,\mathrm{grad}=\nabla_{l}h(l),\mathrm{hessian}=\nabla_{l}^{2}h(l)\Bigg)$
\item $\bar{p} = p(\theta)$
\end{enumerate}
\end{enumerate}}
\EndFunction
\end{algorithmic} 
\caption{Minimum Relative Entropy} \label{alg:mre}
\end{algorithm}
\FloatBarrier

\bibliographystyle{acm}
\bibliography{A2010M}

\begin{thebibliography}{10}

\bibitem{R2016}
{\sc Bailey, D., Borwein, J., Lopez~de Prado, M., and Zhu, Q.}
\newblock The probability of backtest overfitting.
\newblock {\em Journal of Computational Finance 20\/} (04 2017), 39--69.

\bibitem{R2012}
{\sc Bailey, D., and Lopez~de Prado, M.}
\newblock The sharpe ratio efficient frontier.
\newblock {\em The Journal of Risk 15\/} (12 2012), 3--44.

\bibitem{RR2014}
{\sc Bailey, D., and Lopez~de Prado, M.}
\newblock The deflated sharpe ratio: Correcting for selection bias, backtest
  overfitting, and non-normality.
\newblock {\em The Journal of Portfolio Management 40\/} (09 2014), 94--107.

\bibitem{BM2001}
{\sc Benartzi, S., and Thaler, R.}
\newblock Naive diversification strategics in retirement saving plans.
\newblock {\em Amer. Econ. Rev 91\/} (01 2001), 78--98.

\bibitem{C1943}
{\sc Bhattacharyya, A.}
\newblock On a measure of divergence between two statistical populations
  defined by their probability distributions.
\newblock {\em Bull. Calc. Math. Soc. 35\/} (11 1942).

\bibitem{Z2006}
{\sc Boucher, C.}
\newblock Stock prices, inflation and stock returns predictability.
\newblock {\em SSRN Electronic Journal 27\/} (12 2004).

\bibitem{A2000}
{\sc Campbell, J.}
\newblock Asset pricing at the millennium.
\newblock {\em Journal of Finance 55\/} (02 2000), 1515--1567.

\bibitem{Z2114}
{\sc Christiansen, C., Nygaard~Eriksen, J., and Vinther~Møller, S.}
\newblock Forecasting us recessions: The role of sentiment.
\newblock {\em Journal of Banking and Finance 49\/} (12 2014), 459--468.

\bibitem{C2000}
{\sc Connor, G.}
\newblock Book review. active portfolio management: A quantitative approach to
  providing superior returns and controlling risk, 2nd edition. rc grinold, rn
  kahn.
\newblock {\em Review of Financial Studies - REV FINANC STUD 13\/} (10 2000),
  1153--1156.

\bibitem{B1983}
{\sc D.~Arnott, R., and N.~Von~Germeten, J.}
\newblock Systematic asset allocation.
\newblock {\em Financial Analysts Journal - FINANC ANAL J 39\/} (11 1983),
  31--38.

\bibitem{R2018}
{\sc D.~Arnott, R., R.~Harvey, C., and Markowitz, H.}
\newblock A backtesting protocol in the era of machine learning.
\newblock {\em SSRN Electronic Journal\/} (01 2018).

\bibitem{Z2002}
{\sc Darrat, A., and Zhong, M.}
\newblock Permanent and transitory driving forces in the asian‐pacific stock
  markets.
\newblock {\em Financial Review 37\/} (02 2002), 35 -- 51.

\bibitem{Z1993}
{\sc Dasgupta, S., and Lahiri, K.}
\newblock On the use of dispersion measures from napm surveys in business cycle
  forecasting.
\newblock {\em Journal of Forecasting 12\/} (04 1993), 239 -- 253.

\bibitem{BM2009}
{\sc Demiguel, V., Garlappi, L., and Uppal, R.}
\newblock Optimal versus naive diversification: How inefficient is the 1/n
  portfolio strategy?
\newblock {\em Review of Financial Studies 22\/} (05 2009).

\bibitem{Z2015}
{\sc Dreger, C., and Schumacher, C.}
\newblock Out-of-sample performance of leading indicators for the german
  business cycle: Single vs. combined forecasts.
\newblock {\em Journal of Business Cycle Measurement and Analysis 2005\/} (06
  2005), 3--3.

\bibitem{A2011}
{\sc Eychenne, K., and Roncalli, T.}
\newblock Strategic asset allocation.
\newblock {\em SSRN Electronic Journal\/} (03 2011).

\bibitem{A2003}
{\sc Flavin, T., and Wickens, M.}
\newblock Macroeconomic influences on optimal asset allocation.
\newblock {\em Review of Financial Economics 12\/} (02 2003), 207--231.

\bibitem{BM2003}
{\sc Goyal, A., and Welch, I.}
\newblock Predicting the equity premium with dividend ratios.
\newblock {\em Management Science 49\/} (03 2002).

\bibitem{A2007}
{\sc Guidolin, M., and Timmermann, A.}
\newblock Asset allocation under multivariate regime switching.
\newblock {\em Journal of Economic Dynamics and Control 31\/} (01 2008),
  3503--3544.

\bibitem{Z2111}
{\sc Gupta, R., and Modise, M.}
\newblock Macroeconomic variables and south african stock return
  predictability.
\newblock {\em Economic Modelling 30\/} (04 2011).

\bibitem{C1909}
{\sc Hellinger, E.}
\newblock Neue begr??ndung der theorie quadratischer formen von unendlichvielen
  ver??nderlichen.
\newblock {\em Journal Fur Die Reine Und Angewandte Mathematik - J REINE ANGEW
  MATH 1909\/} (01 1909), 210--271.

\bibitem{Z2011}
{\sc Hsing, Y.}
\newblock The stock market and macroeconomic variables in a brics country and
  policy implications.
\newblock {\em International Journal of Economics and Financial Issues 1\/} (01
  2011), 12--18.

\bibitem{Z2014}
{\sc Huang, D., Jiang, F., Tu, J., and Zhou, G.}
\newblock Forecasting stock returns in good and bad times: The role of market
  states.
\newblock {\em 27th Australasian Finance and Banking Conference 2014 Paper;
  Asian Finance Association (AsianFA) 2016 Conference\/} (07 2017).

\bibitem{Z2003}
{\sc Jostova, G.}
\newblock Predictability in emerging sovereign debt markets.
\newblock {\em The Journal of Business 79\/} (03 2006), 527--566.

\bibitem{E2013}
{\sc Kollár, M.}
\newblock A sketch of macro based asset allocation.
\newblock {\em International Journal of Economic Sciences 2(3)\/} (2013),
  101–120.

\bibitem{B1951}
{\sc Kullback, S., and A.~Leibler, R.}
\newblock On information and sufficiency.
\newblock {\em Annals of Mathematical Statistics 22\/} (03 1951), 79--86.

\bibitem{BM1999}
{\sc L.~Bossaerts, P., and Hillion, P.}
\newblock Implementing statistical criteria to select return forecasting
  models: What do we learn?
\newblock {\em Review of Financial Studies 12\/} (02 1999), 405--28.

\bibitem{AA2019}
{\sc Lopez~de Prado, M., and J.~Lewis, M.}
\newblock Detection of false investment strategies using unsupervised learning
  methods.
\newblock {\em Quantitative Finance 19\/} (07 2019), 1--11.

\bibitem{B1969}
{\sc M.~Bates, J., and W.J.~Granger, C.}
\newblock The combination of forecasts.
\newblock {\em OR 20\/} (12 1969), 451--468.

\bibitem{A1952}
{\sc Markowitz, H.}
\newblock Portfolio selection.
\newblock {\em The journal of finance 7(1)\/} (03 1952), 77–91.

\bibitem{B2008}
{\sc Meucci, A.}
\newblock Fully flexible views: Theory and practice.
\newblock {\em Risk Magazine 21\/} (2008).

\bibitem{A2010}
{\sc Meucci, A.}
\newblock Historical scenarios with fully flexible probabilities.
\newblock {\em SSRN Electronic Journal\/} (10 2010).

\bibitem{A2012}
{\sc Meucci, A.}
\newblock Effective number of scenarios in fully flexible probabilities.

\bibitem{A2013}
{\sc Meucci, A.}
\newblock Estimation and stress-testing via time- and market-conditional
  flexible probabilities.

\bibitem{Z2008}
{\sc Munro, B., and Silberman, K.}
\newblock Optimal asset allocation in different economic environments.
\newblock {\em Cadiz Securities Quantitative Research Report\/} (2008).

\bibitem{Z1991}
{\sc Nazmi, N.}
\newblock Leading economic indicators: New approaches and forecasting record :
  Review of lahiri k. and g. moore (eds.), 1991, (cambridge university press,
  cambridge).
\newblock {\em International Journal of Forecasting 10\/} (02 1994), 382--385.

\bibitem{A1986}
{\sc P.~Brinson, G., Randolph~Hood, L., and L.~Beebower, G.}
\newblock Determinants of portfolio performance.
\newblock {\em Financial Analysts Journal - FINANC ANAL J 42\/} (07 1986),
  39--44.

\bibitem{A1998}
{\sc P.~Richardson, M., Boudoukh, J., and F.~Whitelaw, R.}
\newblock The best of both worlds: A hybrid approach to calculating value at
  risk.
\newblock {\em RISK 11\/} (11 1997).

\bibitem{Z2012}
{\sc (Paul)~Dou, Y., Gallagher, D., H.~Schneider, D., and Walter, T.}
\newblock Out-of-sample stock return predictability in australia.
\newblock {\em Australian Journal of Management 37\/} (12 2012), 461--479.

\bibitem{Z2018}
{\sc Qadan, M., Kliger, D., and Chen, N.}
\newblock Idiosyncratic volatility, the vix and stock returns.
\newblock {\em The North American Journal of Economics and Finance 47\/} (06
  2018).

\bibitem{C2010}
{\sc Rapach, D., Strauss, J., and Zhou, G.}
\newblock Out-of-sample equity premium prediction: Combination forecasts and
  links to the real economy.
\newblock {\em Review of Financial Studies 23\/} (04 2009), 821--862.

\bibitem{Z2112}
{\sc Rapach, D., Strauss, J., and Zhou, G.}
\newblock International stock return predictability: What is the role of the
  united states?
\newblock {\em Capital Markets: Market Efficiency eJournal 68\/} (03 2010).

\bibitem{B2013}
{\sc Rapach, D., and Zhou, G.}
\newblock Forecasting stock returns.
\newblock {\em Handbook of Economic Forecasting 2\/} (12 2013), 327--383.

\bibitem{E1994}
{\sc Sharpe, W.}
\newblock The sharpe ratio.
\newblock {\em J Portf Manag 21\/} (01 1994), 49--58.

\bibitem{C1958}
{\sc Sokal, R., and Michener, C.}
\newblock A statistical method of evaluating systematic relationships.
\newblock {\em The University of Kansas Science Bulletin 38\/} (01 1958),
  1409--1438.

\bibitem{C2006}
{\sc Timmermann, A.}
\newblock Chapter 4 forecast combinations.
\newblock {\em Handbook of Economic Forecasting 1\/} (12 2006), 135--196.

\bibitem{Z2093}
{\sc Titman, S., and Jegadeesh, N.}
\newblock Returns to buying winners and selling losers: Implications for stock
  market efficiency.
\newblock {\em Journal of Finance 48\/} (02 1993), 65--91.

\bibitem{Vap1971}
{\sc Vapnik, V., and Chervonenkis, A.}
\newblock On the uniform convergence of relative frequencies of events to their
  probabilities.
\newblock {\em Theory of Probability and Its Applications 16\/} (01 1971),
  264--280.

\bibitem{Z2005}
{\sc Venter, J.}
\newblock A brief history of business cycle analysis in south africa.

\bibitem{B2000}
{\sc Wai, L.}
\newblock Theory and methodology of tactical asset allocation.
\newblock {\em The Frank J. Fabozzi Series\/} (08 2000).

\bibitem{A2008}
{\sc Welch, I., and Goyal, A.}
\newblock A comprehensive look at the empirical performance of equity premium
  prediction.
\newblock {\em Review of Financial Studies 21\/} (02 2008), 1455--1508.

\bibitem{BM2004}
{\sc Windcli, H., and Boyle, P.}
\newblock The 1/n pension investment puzzle.
\newblock {\em N. Am. Actuar. J. 8\/} (12 2003).

\bibitem{Z2000}
{\sc Wu, Y.}
\newblock Stock prices and exchange rates in vec model—the case of singapore
  in the 1990s.
\newblock {\em Journal of Economics and Finance 24\/} (04 2012), 260--274.

\end{thebibliography}

\end{document}